\documentclass[acmsmall]{acmart}
\acmJournal{TOSEM}

\setcopyright{rightsretained}
\copyrightyear{2024}           
\usepackage{graphicx}
\usepackage{subcaption}
\usepackage{threeparttable}
\usepackage{pifont}

\usepackage{tabularx,booktabs,xcolor,colortbl,tcolorbox}
\usepackage{subcaption}
\usepackage[ruled,vlined]{algorithm2e}

\usepackage{amsthm}
\usepackage{graphicx}
\usepackage{amsmath}
\usepackage{mathrsfs}
\usepackage{stmaryrd}
\usepackage{scalerel}
\usepackage{listings}
\usepackage{rotating} 
\usepackage{multirow}
\usepackage{booktabs}
\usepackage{enumitem}

\usepackage{tikz}

\usepackage{etoolbox}

\usepackage{hyperref} \hypersetup{hidelinks,
	colorlinks=true,
	allcolors=black,
	pdfstartview=Fit,
	breaklinks=true
}
\usepackage{bookmark}

\usepackage{pifont}

\usepackage[bbgreekl]{mathbbol}
\DeclareSymbolFontAlphabet{\mathbbm}{bbold}
\DeclareSymbolFontAlphabet{\mathbb}{AMSb}\usetikzlibrary{shapes,arrows,positioning,shadows}
\definecolor{hgreen}{HTML}{8CB26E}
\definecolor{mgreen}{HTML}{A9C695}
\definecolor{lgreen}{HTML}{D9E7D6}
\definecolor{hred}{HTML}{AC5A54}
\definecolor{mred}{HTML}{DCADA9}
\definecolor{lred}{HTML}{F1D0CD}
\definecolor{hblue}{HTML}{738DBB}
\definecolor{mblue}{HTML}{98ADD1}
\definecolor{lblue}{HTML}{DDE8FA}
\definecolor{hgray}{HTML}{585858}
\definecolor{mgray}{HTML}{BCBCBC}
\definecolor{lgray}{HTML}{F0F0F0}
\definecolor{hyel}{HTML}{D1B765}
\definecolor{myel}{HTML}{EEDDAA}
\definecolor{lyel}{HTML}{FDF2D0}
\definecolor{hpur}{HTML}{9174A3}
\definecolor{mpur}{HTML}{C4B3CE}
\definecolor{lpur}{HTML}{DFD5E6}
\definecolor{myhgreen}{RGB}{0,102,0}
\definecolor{spoqLBLUE}{HTML}{D5EBF5}
\definecolor{spoqYELLOW}{HTML}{FFF7A2}
\definecolor{spoqBLUE}{HTML}{65A7D3}
\definecolor{spoqGREEN}{HTML}{BCE1AE}
\definecolor{dblue}{RGB}{0, 0, 139}

\definecolor{mydarkgray}{RGB}{64,64,64}
\definecolor{mylightgray}{RGB}{192,192,192}

\newif\ifshowcomments
\showcommentstrue

\ifshowcomments
\newcommand{\yao}[1]{\mytodoblue{[yao: #1]}}
\newcommand{\hong}[1]{\mytodoorange{[hong: #1]}}
\newcommand{\zeng}[1]{\mytodored{[zeng: #1]}}
\else
\newcommand{\yao}[1]{}
\newcommand{\hong}[1]{}
\newcommand{\zeng}[1]{}
\fi

\newcommand{\mytodoblue}[1]{\textcolor{blue}{\ding{46}~{\sf}~#1}}
\newcommand{\mytodored}[1]{\textcolor{red}{\ding{46}~{\sf}~#1}}

\newcommand{\mytodoorange}[1]{\textcolor{orange}{\ding{46}~{\sf}~#1}}

\newcolumntype{C}{>{\centering\arraybackslash}X}

\newcommand{\hForm}[1]{\textcolor[RGB]{78,137,136}{\ensuremath{#1}}}

\newcommand{\red}[1]{\textcolor{red}{#1}}
\newcommand{\blue}[1]{\textcolor{blue}{#1}}

\newenvironment{lstcenter}[1][0.8\linewidth]
  {\begin{center}\begin{minipage}{#1}}
  {\end{minipage}\end{center}}
\usepackage{color, xcolor} 

\definecolor{kterminalred}{RGB}{188,46,47}
\definecolor{knonterminalpurple}{RGB}{70, 32, 126}
\definecolor{ksyntaxgreen}{RGB}{1,86,153}
\definecolor{ksyntaxblack}{RGB}{2,15,41}
\definecolor{kcomments}{RGB}{78,137,136}
\definecolor{codegray}{rgb}{0.5,0.5,0.5}

\usepackage{listings} 
\lstdefinelanguage{K}{
	upquote=true,
	sensitive=false,
	morekeywords={syntax, configuration, rule},
	keywordstyle=\bfseries,
	comment=[l][\color{kcomments}]{//}, 	morestring=[b]",
	stringstyle=\color{gray},
    literate={"}{\textquotedbl}1,
		}

\newcommand{\inlineK}[1]{``\lstinline[language=K]{#1}''}
\newcommand{\inlineKnoQ}[1]{\lstinline[language=K]{#1}}

\newcommand{\Pname}{Formal Synchronizer Synthesis}
\newcommand{\Pid}{FSS}
\newcommand{\K}{$\mathbb{K}$}
\newcommand{\SaC}{\small\colorbox{spoqLBLUE}{State+C}}
\newcommand{\putr}{\small\colorbox{spoqBLUE}{\color{white}PutR}}
\newcommand{\creater}{\small\colorbox{spoqBLUE}{\color{white}CreateR}}
\newcommand{\defaultV}{\small\colorbox{spoqYELLOW}{Default Value}}

\usepackage{xcolor} 
\newif\ifshowtodos
\showtodosfalse

\definecolor{turquoise}{RGB}{64,224,208}
\definecolor{darkteal}{RGB}{0,128,128}

\ifshowtodos
\newcommand{\ToResearch}[1]{\textcolor[RGB]{255,0,0}{[To Research: #1]}} \newcommand{\ToCite}[1]{\textcolor[RGB]{0,128,0}{[To Cite: #1]}} \newcommand{\ToRead}[1]{\textcolor[RGB]{0,0,255}{[To Read: #1]}} \newcommand{\ToWrite}[1]{\textcolor[RGB]{128,0,128}{[To Write: #1]}} \newcommand{\ToPreWork}[1]{\textcolor[RGB]{255,165,0}{[To PreWork: #1]}} \newcommand{\ToOpt}[1]{\textcolor[RGB]{255,20,147}{[To Opt: #1]}} \newcommand{\ToStartMyDay}[1]{\colorbox{darkteal}{\textcolor{white}{[To Start My Day: #1]}}}
\newcommand{\DateMark}[2]{\textcolor[RGB]{139,69,19}{[#1: #2]}} \newcommand{\ToAddInfo}[1]{\textcolor[RGB]{0,255,255}{[To Add Info: #1]}} \newcommand{\ToCheck}[1]{\textcolor[RGB]{255,191,0}{[To Check: #1]}} \newcommand{\ToDelete}[1]{\textcolor[RGB]{105,105,105}{[ToDelete: #1]}} \else
\newcommand{\ToResearch}[1]{}
\newcommand{\ToCite}[1]{}
\newcommand{\ToRead}[1]{}
\newcommand{\ToWrite}[1]{}
\newcommand{\ToPreWork}[1]{}
\newcommand{\ToOpt}[1]{}
\newcommand{\ToStartMyDay}[1]{}
\newcommand{\DateMark}[2]{}
\newcommand{\ToAddInfo}[1]{}
\newcommand{\ToCheck}[1]{}
\newcommand{\ToDelete}[1]{}
\fi

\begin{document}

\title{KBX: Verified Model Synchronization via Formal Bidirectional Transformation}

\thanks{This work has been supported by the
\grantsponsor{}{Natural Science Foundation of China}{} under Grants
No.~\grantnum{}{U2341212} and
No.~\grantnum{}{62132014}.}

\author{Jianhong Zhao}
\email{zhaojianhong96@gmail.com}

\author{Yongwang Zhao}
\authornote{Corresponding author}
\email{zhaoyw@zju.edu.cn}

\author{Peisen Yao}
\email{pyaoaa@zju.edu.cn}

\author{Fanlang Zeng}
\email{flzeng@zju.edu.cn}
\affiliation{  \department[0]{School of Cyber Science and Technology}
  \department[1]{College of Computer Science and Technology}
    \institution{Zhejiang University}
    \city{Hangzhou}
  \state{Zhejiang}
  \country{China}
  }

\author{Bohua Zhan}
\email{bzhan@ios.ac.cn}
\affiliation{  \institution{Institute of Software Chinese Academy of Sciences}
    \city{Beijing}
    \country{China}
  }

\author{Kui Ren}
\email{kuiren@zju.edu.cn}
\affiliation{  \department{College of Computer Science and Technology}
  \institution{Zhejiang University}
    \city{Hangzhou}
  \state{Zhejiang}
  \country{China}
  }
\thanks{Y. Zhao, P. Yao, and K. Ren also work at the State Key Laboratory of Blockchain and Data Security, Zhejiang University.}

\renewcommand{\shortauthors}{Zhao et al.}

\begin{abstract}
Complex safety-critical systems require multiple models for a comprehensive description, resulting in error-prone development and laborious verification. Bidirectional transformation (BX) is an approach to automatically synchronizing these models. However, existing BX frameworks lack formal verification to enforce these models' consistency rigorously.
This paper introduces KBX, a formal bidirectional transformation framework for verified model synchronization. First, we present a matching logic-based BX model, providing a logical foundation for constructing BX definitions within the $\mathbb{K}$ framework.
Second, we propose algorithms to synthesize formal BX definitions from unidirectional ones, which allows developers to focus on crafting the unidirectional definitions while disregarding the reverse direction and missing information recovery for synchronization. Afterward, we harness $\mathbb{K}$ to generate a formal synchronizer from the synthesized definitions for consistency maintenance and verification. 
To evaluate the effectiveness of KBX, we conduct a comparative analysis against existing BX frameworks. Furthermore, we demonstrate the application of KBX in constructing a BX between UML and HCSP for real-world scenarios, showcasing an 82.8\% reduction in BX development effort compared to manual specification writing in $\mathbb{K}$.

\end{abstract}

\ccsdesc[500]{Computing methodologies~Modeling methodologies}
\ccsdesc[500]{Software and its engineering~Formal software verification}

\keywords{Bidirectional Transformation, Formal Verification, Matching Logic}

\maketitle

\section{Introduction}
\label{sec:intro}

Modeling and verifying safety-critical systems~\cite{kulikSurveyPracticalFormal2022} can be a complex task that involves the use of diverse languages and abstraction levels.
This complexity arises from the limitations of single languages, the complexity of systems, and the diversified requirement of different standards (e.g., ISO-15408 \cite{CommonCriteriaInformation2017b} and IEC-61580 \cite{bellIntroductionIEC615082006}). For instance, seL4 \cite{kleinRefinementFormalVerification2010a,kleinComprehensiveFormalVerification2014} proposes abstract specification in Isabelle, executable prototype in Haskell, and high-performance implementation in C. To obtain a verified system, it uses refinement verification and unidirectional transformers from programming languages (Haskell and C) to Isabelle.

Unfortunately, applying multiple models needs extra effort to synchronize models and verify their consistency. Any modification in one model (Isabelle specification) mandates manual adjustments in others (Haskell prototype and C implementation). To satisfy the stringent traceability requirements of safety-critical systems, this error-prone process is vital for proving the consistency of models post their transformation to a formal framework like Isabelle. The process becomes increasingly cumbersome, especially for developers not proficient in multiple languages, and is exacerbated by the growing number of models and iterations involved.

To tackle this problem, our vision is a formal bidirectional transformation (BX) framework. This framework produces synchronizers for \textit{model synchronization}, and formal proofs for rigorous \textit{consistency verification} of these models.
In this way, the developers can credibly focus on their familiar models while the formal BX automatically enforces consistency. However, verified model synchronization in diverse languages and abstract levels via BX is stunningly challenging, considering the following important criteria:

\smallskip
\textbf{1. Expressiveness}.
Complicate synchronizations necessitate expressiveness to specify intricate consistency between heterogeneous models. To achieve this, BX frameworks should help developers focus on consistency without being burdened by extraneous concepts or implementation details. Moreover, there should be ample evidence demonstrating that developers can effectively construct diverse transformations using BX frameworks. However, the existing BX frameworks \cite{anjorinBenchmarkingBidirectionalTransformations2020,buchmannBXtendDSLLayeredFramework2022,buchmannBXtendAFrameworkBidirectional2018,koBiGULFormallyVerified2016,bettiniImplementingDomainspecificLanguages2016,cicchettiJTLBidirectionalChange2011,weidmannIncrementalBidirectionalModel2019b,matsudaHobitProgrammingLenses2018} require provided manually parsers and printers, introducing extra implementation work and concept of abstract syntax trees.
Additionally, few evidence shows that they can describe sophisticated model BX (e.g., UML and concurrent HCSP\footnote{HCSP (hybrid communicating sequential processes) \cite{chaochenFormalDescriptionHybrid1996,liuCalculusHybridCSP2010} is an extension of CSP with ordinary differential equations, and it is widely used in verifying cyber-physical systems such as train control systems \cite{zouVerifyingSimulinkDiagrams2013,zouFormalVerificationSimulink2015}, cruise control system \cite{xuUnifiedGraphicalComodeling2022}, and Mars lander \cite{zhanBriefIndustryPaper2021}.
} BX) in practice. 

\smallskip
\textbf{2. Trustworthiness}. 
Rigorous verification for safety-critical systems demands a formal method with a small trust base. However, existing BX frameworks lack a formal proof system for verification. Moreover, the existing formal frameworks such as Coq~\cite{WelcomeCoqProof}, Isabelle~\cite{Isabelle}, Lean~\cite{Lean}, and $\mathbb{K}$~\cite{FrameworkTools2023} offer rigorous formal verification but lack solutions for verified model synchronization. First, they require a uniform BX model that can capture a reasonable BX within their theories. Second, they need synthesis algorithms to provide a minimized formal specification as a small trust base to generate a formal synchronizer. This synchronizer not only achieves rationale synchronization, as existing BX frameworks do, but also offers synchronization that adheres to a formal definition, supported by proofs.

Our key idea comes from the work of \citet{chenTrustworthySemanticsBasedLanguage2021}, which introduced the formal verification of the program execution via proof generation.
By applying this approach, the $\mathbb{K}$ framework can automatically verify that the transformation complies with the established formal transformation definition, using the matching logic proof system. In addition to its verification capabilities, the $\mathbb{K}$ framework possesses a \textit{language-oriented} nature, enabling the construction of intricate formal semantics intuitively. This feature has led to numerous programming languages, including C \cite{hathhornDefiningUndefinedness2015}, Java \cite{bogdanasKJavaCompleteSemantics2015}, and JavaScript \cite{parkKJSCompleteFormal2015}, having their complete formal semantics defined in the $\mathbb{K}$ framework, rather than just partial semantics in other formal frameworks. Consequently, we can construct formal unidirectional transformation definitions within $\mathbb{K}$ to generate a verified transformer for model transformation while ensuring compliance with the formal definition.

Nevertheless, formal unidirectional transformation cannot deliver missing information recovery and reverse transformation, as offered by bidirectional transformation for synchronization. Therefore, this paper introduces KBX, an extension to the $\mathbb{K}$ framework, enabling formal BX for verified model synchronization. To achieve this, we employ three steps: (1) Capture BX models using matching logic from $\mathbb{K}$ to facilitate the construction of formal BX $\mathbb{K}$ definitions. (2) Design formal BX definition synthesis algorithms to generate formal synchronizers from unidirectional transformation definitions. The BX definition incorporats both forward and backward transformation definitions to enable missing information recovery for synchronized target. (3) Use formal synchronizers to maintain and verify the consistency of models simultaneously. 

We initiate our evaluation of KBX by comparing its development efficiency and synchronization trustworthiness with current BX frameworks. This analysis highlights KBX's efficacy in constructing trustworthy BX programs. Next, we explore KBX's formal expressiveness and verification reliability, demonstrating that we impose limited constraints on the expressiveness of $\mathbb{K}$ and maintain a small trust base during verification. Lastly, we illustrate KBX's practicality in real-world scenarios by establishing the first formal BX between HCSP and UML. During the construction of this BX, KBX reduces 82.8\% code size, compared with manual writing BX definitions.

To summarize, this paper makes the following main contributions:

\begin{itemize}
  \item  We present KBX, the BX framework for verified model synchronization, which generates formal synchronizers from formal unidirectional transformation definitions.
  \item We introduce a matching logic-based BX model to establish the relation between unidirectional definitions and BX definitions, as well as to specify the laws of BX definitions for synchronization. Furthermore, we present BX synthesis algorithms aimed at improving verification trustworthiness and accelerating BX development. Based on the synthesis algorithms, we generate formal BX definitions from unidirectional ones in $\mathbb{K}$, producing formal synchronizers for simultaneous synchronization and verification.
  \item We evaluate KBX's efficacy in synchronization and verification. Additionally, we propose the first HCSP and UML BX for cyber-physical systems based on KBX, demonstrating its practical applicability in real-world scenarios.
  \end{itemize}

\section{Preliminaries}

This section introduces the foundations of this work, including bidirectional transformation, matching logic, and the $\mathbb{K}$ framework.

\subsection{Bidirectional Transformation} \label{subsec:symmetric-lenses}

Bidirectional transformation (BX) enables the conversion of a system between two representations and maintains the synchronization of changes between them. For instance, it enables the synchronization of system designs between HCSP and plantUML by transforming one from another while keeping the missing information unchanged. A prominent theory elucidating bidirectional transformation is referred to as symmetric lenses~\cite{hofmannSymmetricLenses2011}, which permits information loss on both ends of such transformations. The \textit{symmetric lens} is defined as follows.

\begin{definition}[Symmetric Lens]\label{def:symmetric-lens} Lens $\ell$ of model sets $M$, $N$ (denoted as $\ell \in M \leftrightarrow N$) has three parts: a set of complements $C$, a distinguished element $missing \in C$, and two functions
  $$ putr\ :\ M \times C \rightarrow N \times C \ \ \ \ \ \  \ putl\ :\ N \times C \rightarrow M \times C$$
  which satisfy the following round-tripping laws:
  \begin{align*}
    \begin{tabular}{cc}
      $putr(m,c)$          & $= (n,c^{\prime})$  \\ \cline{1-2}
      $putl(n,c^{\prime})$ & $= (m, c^{\prime})$ \\
    \end{tabular}\label{def:oputrl}\tag{P{\scriptsize UT}RL} \\
    \begin{tabular}{cc}
      $putl(n,c)$          & $= (m,c^{\prime})$ \\ \cline{1-2}
      $putr(m,c^{\prime})$ & $= (n,c^{\prime})$ \\
    \end{tabular}\label{def:oputlr}\tag{P{\scriptsize UT}LR}
  \end{align*}
\end{definition}

The $missing$ element in complements $C$ represents missing information to be recovered for synchronization. For example, when synchronizing from HCSP to UML, we expect to update the modifications and retain the rest, including the missing information, in the UML model.
The primitive unidirectional transformation ($M \rightarrow N$) can maintain the shared information but cannot recover the missing information, such as line colors in UML. In contrast, the forward transformation $putr$ can recover the missing information (i.e., line colors in UML) from the complement $C$ stored by the backward transformation $putl$, thus providing reasonable synchronization. The round-tripping laws formulate this synchronization rationale for forward and backward transformations.

\subsection[Matching Logic and K Framework]{Matching Logic and $\mathbb{K}$ Framework} 
\label{subsec:matching-logic}

Matching logic~\cite{chenTrustworthySemanticsBasedLanguage2021,chenMatchingMuLogicFoundation2019,chenMatchingLogicExplained2021} serves as the logical foundation of $\mathbb{K}$ language framework, enabling the specification and reasoning of programs. The simplicity and expressiveness of matching logic enable $\mathbb{K}$ to offer a concise and intuitive approach for defining complex semantics, with practical applications in programming languages such as C~\cite{ellisonExecutableFormalSemantics2011,hathhornDefiningUndefinedness2015}, Java~\cite{bogdanasKJavaCompleteSemantics2015}, and Javascript~\cite{parkKJSCompleteFormal2015}. In Section~\ref{sec:model}, we propose a novel symmetric lens using matching logic that can be programmed in $\mathbb{K}$. Other sections in Section~\ref{sec:approach} further illustrate how to verify and synthesize our lenses in $\mathbb{K}$ based on matching logic.

Matching logic formulas, called \textit{patterns}, serve as unified structures for syntax and semantic specifications. They are inductively defined based on the matching logic signature $\mathbbm{\Sigma}=(EV, SV, \Sigma)$, where $EV$ represents the set of element variables (denoted x, y, ...), $SV$ denotes the set of set variables (denoted X, Y, ...), and $\Sigma$ refers to the set of constant symbols (denoted $\sigma$, $\sigma_1$, ...).

\begin{definition}[Matching logic syntax] \label{def:ml-syntax}
  The set P{\scriptsize ATTERN}($\mathbbm{\Sigma}$) of $\mathbbm{\Sigma}$-patterns is inductively defined as follows:
  $$
    \varphi::=x\ |\ X\ |\ \sigma\ |\ \varphi_{1} \varphi_{2} \ |\ \perp \ |\varphi_{1} \rightarrow \varphi_{2}\ |\ \exists x . \varphi\ |\ \mu X . \varphi
  $$
  where $\varphi$ has no negative occurrences of X in $\mu X . \varphi$.
\end{definition}

Def.\ref{def:ml-syntax} indicates that \textit{patterns} are constructed with variables ($x$, $X$), constant symbols ($\sigma$), applications to apply the first argument to the second ($\varphi_1 \varphi_2$), standard first-order logic (FOL) constructs ($\perp$, $\rightarrow$, $\exists$), and the least fixpoint construct ($\mu$). Notation $\varphi[\psi/x]$ (resp. $\varphi[\psi/X]$) denotes the capture-free substitution of $\psi$ for $x$ (resp. $X$) in $\varphi$. Other notations relevant to this paper are defined as follows:
\begin{align*}
  & \neg \varphi \equiv \varphi \rightarrow \perp \ \   \top \equiv \neg \perp\ \ &&\varphi_{1} \wedge \varphi_{2} \equiv \neg\left(\neg \varphi_{1} \vee \neg \varphi_{2}\right) \\
  & \ \ \ \ \ \ \ \ \ \forall x . \varphi \equiv \neg  \exists x . \neg \varphi  &&\varphi_{1} \vee \varphi_{2} \equiv \neg \varphi_{1} \rightarrow \varphi_{2}
\end{align*}

Matching logic employs a \textit{pattern matching semantics}~\cite{chenMatchingLogicExplained2021}, interpreting a pattern $\varphi$ as a set of elements that match it. For example, $\perp$ corresponds to $\emptyset$, while $\varphi_1 \land \varphi_2$ is matched by elements that match both $\varphi_1$ and $\varphi_2$.

The \textit{matching logic proof system}~\cite{chenTrustworthySemanticsBasedLanguage2021} defines a provability relation, written $\Gamma \vdash \varphi$, meaning that $\varphi$ can be proved using the proof system, with patterns in semantics $\Gamma$ added as additional axioms. We also call $\Gamma$ a \textit{matching logic theory}. The proof system includes the FOL rules for FOL reasoning, frame rules for application context reasoning, fixpoint rules for standard fixpoint reasoning as in modal $\mu$-calculus~\cite{kozenResultsPropositionalMcalculus1983}, and technical proof rules for some completeness results~\cite{chenMatchingMlogic2019}. 

Using matching logic, each $\mathbb{K}$ definition of a language $L$ corresponds to a \textit{matching logic theory} $\Gamma^L \subseteq $ \textit{P{\scriptsize ATTERN}}($\mathbbm{\Sigma}$). This theory essentially comprises a set of logical symbols representing the formal syntax of $L$, and a set of logical axioms specifying the formal semantics. $\mathbb{K}$ generates tools formally specified by matching logic formulas, such as \textit{program execution}, which is represented by the pattern $\varphi_{init} \Rightarrow \varphi_{final}$, where $\varphi_{init}$ and $\varphi_{final}$ denote the initial and final states of execution, respectively. The formal proof $\Gamma^L \vdash \varphi_{init} \Rightarrow \varphi_{final}$ substantiates this pattern.

Program execution can be further delineated into a comprehensive and concrete execution trace $\varphi_0,\varphi_1,...,\varphi_n$, where $\varphi_0 \equiv \varphi_{init}$ and $\varphi_n \equiv \varphi_{final}$. The patterns $\varphi_0,\varphi_1,...,\varphi_n$ represent the intermediate execution \textit{snapshots}. For each step from $\varphi_i$ to $\varphi_{i+1}$, the \textit{rewriting information} includes the applied rewrite/semantic rule $\varphi_{lhs} \Rightarrow \varphi_{rhs}$ and the corresponding substitution $\theta$ such that $\varphi_{lhs}\theta \equiv \varphi_i$.

Specifically, rewriting employs a \textit{one-path next} symbol $\bullet \in \Sigma$ to capture the transition system (or formal semantics) over computation configurations (or states) defined by rewrite rules in $\mathbb{K}$. This symbol denotes that for any configuration $\gamma$, $\bullet \gamma$ is matched by all configurations that can go to $\gamma$ in one step. In other words, $\gamma$ is reached on \textit{one-path} in the \textit{next} configuration. In Def.\ref{def:rewriting}, patterns $\varphi_{1},...,\bullet\bullet\varphi_{2}, \bullet\varphi_{2}, \varphi_{2}$ are execution \textit{snapshots} within an execution trace, where $\varphi_{1}$ denotes the inital state $\varphi_{init}$ and $\varphi_{2}$ sginifies the final state $\varphi_{final}$.

\begin{definition}\label{def:rewriting}
  Program execution (i.e., rewriting) is the reflexive and transitive closure of one-path next, which can be defined as follows:
  \begin{align*}
    \diamond \varphi & \equiv \mu X . \varphi \vee \bullet X\ \text {// Eventually;}\ \varphi \vee \bullet \varphi \vee \bullet \bullet \varphi \vee \ldots \\
    \varphi_{1} \Rightarrow \varphi_{2} & \equiv \varphi_{1} \rightarrow \diamond \varphi_{2} \ \text{// Rewriting }
  \end{align*}
\end{definition}

Leveraging matching logic for formal semantics, the $\mathbb{K}$ framework introduces keywords \textbf{\textit{syntax}} for syntax definition, \textbf{\textit{configuration}} for state declaration, \textbf{\textit{rule}} for executable formal semantics description, and \textbf{\textit{claim}} for constructing a reachability specification under verification. A \textit{rule} consists of a \textit{left-hand side} and a \textit{right-hand side}, separated by the rewrite operator $\Rightarrow$. When the configuration matches the left-hand side of the rule, the rule applies, resulting in the substitution of the configuration with the right-hand side. Additionally, a rule may include a side condition, introduced by the \textit{\textbf{requires}} keyword. The condition is allowed to reference variables that appear on the left-hand side of the rule.

\smallskip
\begin{lstcenter}[0.71\linewidth]
\begin{lstlisting}[language=K, escapeinside={(*}{*)}]
 syntax Weight := Int
 syntax Category := "Light" | "Heavy" 
 configuration <apple> $PGM:K </apple> 
 rule <apple> W:Weight => Light </apple> requires W <Int 100
\end{lstlisting}
\end{lstcenter}

For example, in the provided code snippet, the user can assign a weight to an apple in the cell labeled ``<apple>''. If the weight falls below 100, the content of the cell will display as ``Light''. 
Since the theories link to $\mathbb{K}$ and capture the semantics~\cite{chenTrustworthySemanticsBasedLanguage2021}, we are free to use the theories to capture our theories (or definitions) about symmetric lenses to make them without instantiable in $\mathbb{K}$ considering the semantics in this paper.

\section{Research Overview}

In this section, we first clarify the motivation of our work via an industrial example from high-speed maglev train verification.
We then present an overview of KBX and state the challenges of verified model synchronization.
\subsection{Motivating Example}
\label{sec:motivating-example}
This work is motivated by an industrial practice of verifying 
high-speed maglev trains, which are heterogeneous safety-critical systems with frequent and complex communications. To ensure the intuitive yet rigorous design of such systems, we collaborate with our industrial partners and utilize two distinct models: (1) an abstract model in UML~\cite{PlantUML2022} that is rough but intuitive for team communication; and (2) a refined model in HCSP~\cite{liuCalculusHybridCSP2010}, a formal language for hybrid system simulation and verification.

\begin{figure}[t]
  \centering
  \includegraphics[width=\linewidth]{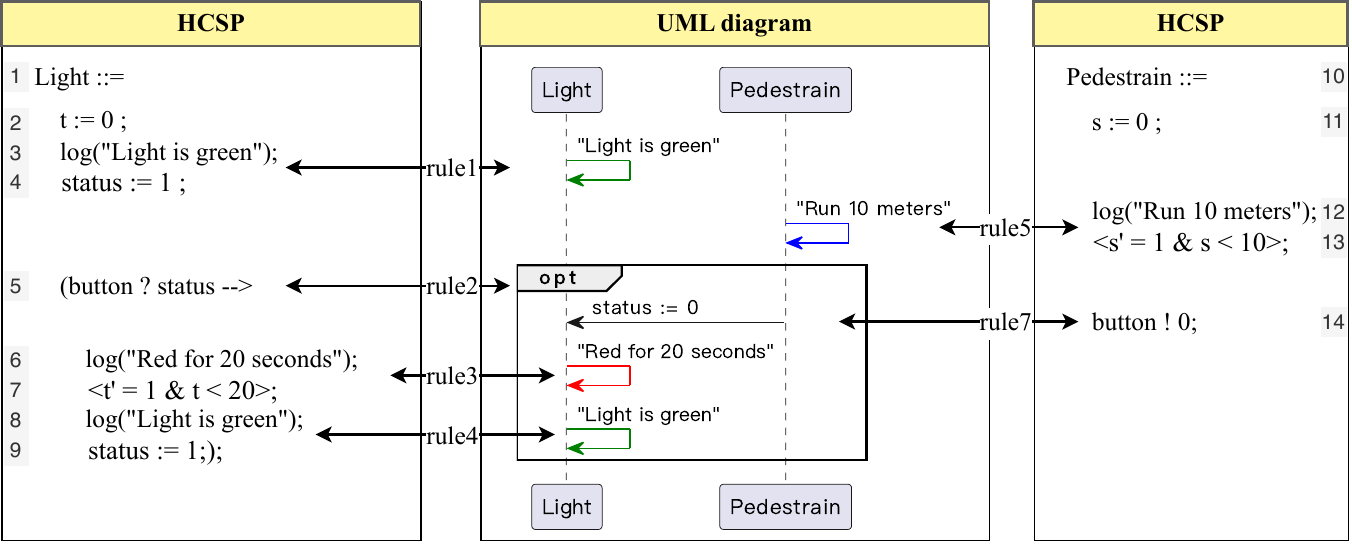}
  \caption{Pedestrian Interaction with Traffic Signals: A depiction of a pedestrian crossing the road with a traffic light initially displaying red, followed by a 10-meter walk, button activation, a 20-second green signal, and subsequent return to red.}
  \label{fig:example}
\end{figure}

For instance, in Fig~\ref{fig:example}, the UML model visualizes how pedestrian-controlled traffic lights work. In this model, arrows pointing to oneself represent actions, like ``$Run\ 10\ meters$'', while arrows pointing to others signify interactions initiated by an entity, as seen in ``$status := 0$''. The ``$opt$'' block indicates that the first message can trigger operations in it.
In contrast, the HCSP model refines the UML model by applying specific rules (e.g., ``$rule1$'' converts the message content of UML ``$Light\ is\ green$'' to the log of HCSP, and introduces an expression to refine this behavior ``$status := 1$'').  This refined model introduces elements like assignments, e.g., ``$status := 1$'',  channels, e.g., ``$button\ ?\ status$'', and ordinary differential equations, e.g., ``$<s' = 1\ \&\ s < 10>$''. By exiling details from abstract models to refined models, this approach improves both clarity and precision in the design process. 

However, this approach suffers from trustworthiness issues.
As an example, suppose that the HCSP model fails to synchronize with the UML according to $rule7$. After the pedestrian presses the button, the light should switch to red, ensuring the pedestrian crosses the road. However, due to the lack of ``$button\ !\ 0$'', the light erroneously remains green, posing a substantial risk of traffic accidents. As intricate and safety-critical systems, Maglev trains could lead to even graver consequences if consistency is compromised. Hence, it becomes imperative to synchronize these models and rigorously verify their consistency.

\smallskip
\textbf{Model Synchronization}. Our first task is to maintain consistency by synchronizing the models after any modification. This involves shared information updates and missing information recovery. For instance, if we adjust the pedestrian distance traveled in HCSP from 10 to 5 (lines 12-13 of Fig~\ref{fig:example}), we need to update the message content to ``$Run\ 5\ meters$'' and recover the message color to be blue according to ``$rule5$''. Manually updating all the modifications in shared information is laborious, while developing separate unidirectional transformers with mutually consistent behavior for synchronization can be difficult and error-prone. Using BX frameworks enables the efficient development of bidirectional transformers that satisfy synchronization rationale (e.g., round-tripping laws) for accurate update and recovery.

\smallskip 
\textbf{Consistency Verification}. Our second task is to prove that the synchronized models are consistent. Since models are not always perfectly equivalent, refinement verification becomes a valuable choice. However, this approach involves significant verification efforts. For example, in the case of proving the refinement relation between HCSP and UML, a multi-step process is required \footnote{These steps represent the unwinding process in the ``Forward simulation \& BX'' approach, as presented in Table~\ref{tab:comparison}. This table offers a comparative analysis of consistency maintenance and verification approaches.}: 

\begin{enumerate}
    \item Construct formal semantics for HCSP and UML within a formal framework.
    \item Develop translators to convert HCSP and UML models into formal representations.
    \item Build formal specifications that define the consistency between HCSP and UML.
    \item Verify that the synchronized models conform to the formal specifications. \end{enumerate}

In summary, employing multiple modeling languages allows us to take the unique strengths of each language, but it also introduces complexities that can hamper the development and verification process, leading to inefficiencies and the risk of errors.

\subsection{KBX Overview}
\label{subsec:overview}
To simplify the development and verification of multiple models, our vision is to design a formal bidirectional transformation (BX) framework. First, by automating the synchronization process, BX eliminates the laborious manual language conversions, such as those between PlantUML and HCSP. Second, formal BX takes a step further by addressing the need for \textit{consistency verification}. This is accomplished by employing formal specifications to establish rigorous consistency and utilizing a proof system to verify this consistency. As a result, the framework can (1) automatically synchronize HCSP and UML models, and (2) formally prove the consistency.

However, existing BX frameworks face limitations in both synchronization and verification.
\begin{itemize}
    \item First, existing BX frameworks fall short in expressiveness and trustworthiness for \textit{synchronization}. Specifically, they lack a focus on being \textit{language-oriented} and \textit{formal}. To illustrate the issue of expressiveness, consider the synchronization of HCSP and UML models. Using current BX frameworks, users not only have to implement BX programs but are also required to develop parsers and printers for both HCSP and UML. These parsers enable the BX program to manipulate the models, while the printers produce HCSP and UML representations instead of abstract syntax trees (AST). A more expressive BX framework should eliminate the need for users to develop parsers and printers. 
    Additionally, in terms of trustworthiness, these frameworks often rely on programming languages (e.g., C\#~\cite{hinkelChangePropagationBidirectionality2019}) or diagram languages (e.g., EVL+Strace~\cite{samimi-dehkordiEVLStraceNovel2018}) to describe model relations, rather than formal languages like $\mathbb{K}$, Coq, and HOL. This can result in undefined behavior of the generated synchronizer. 
    \item Second, existing BX frameworks lack formal languages and proof systems, requiring additional effort to provide \textit{consistency verification}. Without formal languages, we cannot provide verification goals; without proof systems, we lack the mechanisms to prove these goals. To illustrate, these frameworks cannot express consistency definitions (e.g., rules in Fig~\ref{fig:example}) using a formal language, nor can they prove that the synchronized HCSP and UML models adhere to these definitions with proofs. Hence, they demand laborious consistency verification, as discussed in Section~\ref{sec:motivating-example}.
\end{itemize}

\begin{figure}[t]
    \centering
    \includegraphics[width=\linewidth]{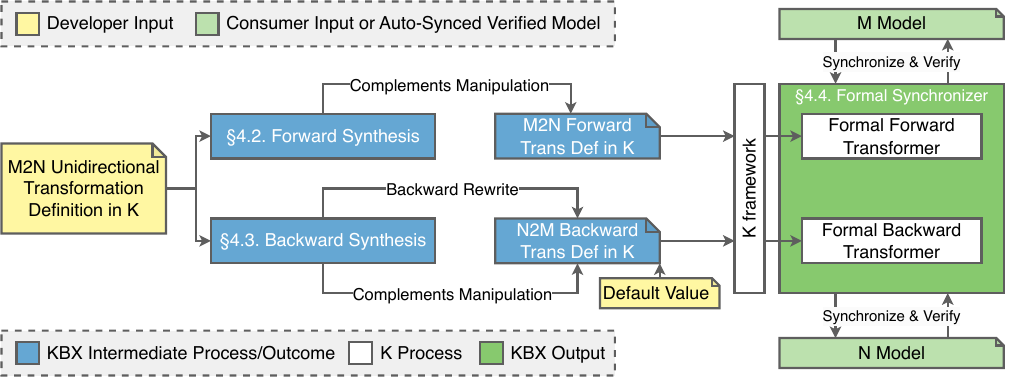}
    \caption{The Architecture of KBX Formal BX Framework.}
    \label{fig:components}
\end{figure}

\smallskip
\textbf{Approach}.
This paper introduces KBX, a novel formal BX framework for simultaneous synchronization and verification, as shown in Fig~\ref{fig:components}. In KBX, developers use unidirectional transformation definitions to generate formal synchronizers, automating model synchronization and consistency verification for consumers. This integration of formal verification not only ensures trustworthiness but also aligns it with high evaluation assurance levels of international safety and security standards, e.g., IEC-61580~\cite{bellIntroductionIEC615082006}, DO-178C~\cite{jacklinCertificationSafetyCriticalSoftware2012}, and ISO-15408~\cite{CommonCriteriaInformation2017b}, making it particularly suitable for safety-critical systems. To elucidate KBX's trustworthiness and expressiveness, the following delves into KBX workflow in Fig.~\ref{fig:components}.

\begin{figure}[ht]
    \centering
    \includegraphics[width=0.7\textwidth]{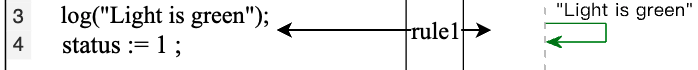}
    \caption{A screenshot of rule1 in Fig.~\ref{fig:example}}
    \label{fig:example-frag}
    \Description{A screenshot of rule1 in Fig.~\ref{fig:example}}
\end{figure}

\paragraph{Expressiveness} KBX streamlines the development of formal synchronizers. Fig.~\ref{fig:uni-trans} illustrates this, showcasing a snippet of developer input, which focuses on formalizing rule 1 (see Fig.~\ref{fig:example-frag}) from HCSP to UML. The syntax for HCSP and UML is defined using the \inlineK{syntax} keyword and BNF grammar in lines 1-3. Specifically, line 2 specifies \inlineK{HCSP} syntax for HCSP programs such as \inlineK{log("Light is green"); status:=1} in Fig.~\ref{fig:example-frag}. The \inlineK{configuration} keyword in line 4 sets up the state's structure, initializing three cells (m, n, and s), with the first cell loaded with an environment variable ``\$PGM'' (the HCSP program). Subsequent lines (5-7), which employ the \inlineK{rule} keyword, formalize the transformation rules from HCSP to UML via matching logic rewriting \inlineK{=>}. For instance, in Fig.~\ref{fig:uni-trans}, \inlineK{status:=1} from Fig~\ref{fig:example-frag} matches with \colorbox{lred}{L}\inlineKnoQ{:=}\colorbox{lred}{R} on the left-hand side, leading to a rewrite on the right-hand side.

Shown in Fig.~\ref{fig:components}, KBX processes these unidirectional transformation definitions to perform its core functions. In the formal synthesis stage (see Section \ref{sec:forward}), KBX enriches each rule with complements manipulations, creating a forward transformation definition that captures and recovers missing information. Conversely, the backward synthesis process (Section \ref{sec:backward}) generates a backward transformation definition in harmony with the forward transformation, ensuring adherence to round-tripping laws. This stage includes the introduction of backward rewrite in matching logic, enhancing the framework's theoretical reliability within the {\K} framework's context. After this stage, developers are responsible for providing default values in the backward transformation when these values are not derivable from the complements. An illustrative example is the necessity to assign default values ``?'' for ambiguous variables \colorbox{lred}{L} and \colorbox{lred}{R} in HCSP assignments, such as \inlineK{? := ?}, since the corresponding information is absent in the UML models. Finally, KBX utilizes the $\mathbb{K}$ framework to automatically generate verifiable and certifiable transformers from these definitions, forming a formal synchronizer for synchronization and verification.

Consequently, KBX's \textit{expressiveness} for developers is manifested in three key aspects: conciseness, intuitiveness, and capability. KBX synthesis processes eliminate the need for developers to grapple with the complexities of synchronization rationale and reverse transformation definitions, contributing to its conciseness. The KBX's intuitive nature, as showcased in lines 5-7 of Fig.~\ref{fig:uni-trans}\footnote{Color boxes in this example highlight matching logic variables for clarity. The symbol \inlineK{[\_]} is used as syntactic sugar for \inlineK{ListItem(\_)}. Being a fragment, this snippet omits certain rules, like those for storing process names and converting programs into \inlineK{List}, typically seen in line 5. Note that this example does not consider the concurrency of HCSP.}, enables developers to seamlessly utilize target syntaxes like HCSP and UML for rule formalization, thus avoiding the intricacies of handling dual ASTs. Lastly, KBX's capability is grounded in the established efficacy of the {\K} framework, which has been successfully applied in the formal semantics of languages including C, Java, and JavaScript.

\begin{figure}[ht]
\begin{lstcenter}[0.72\linewidth]
\begin{lstlisting}[language=K, escapeinside={(*}{*)}]
 syntax HCSPStat ::= "log" "(" String ")" | Id ":=" Expr
 syntax HCSP ::= HCSPStat | HCSPStat ";" HCSP
 syntax UMLStat ::= Id "-[" Color "]>" Id ":" String
 configuration <m> $PGM:HCSP </m> <n> .K </n> <s> .K </s>
 rule <m> [log((*\colorbox{lblue}{A}*)), (*\colorbox{lred}{L}*) := (*\colorbox{lred}{R}*)] (*\colorbox{lgray}{HCSPs}*):List => (*\colorbox{lgray}{HCSPs}*) </m>
      <n> (*\colorbox{lgray}{UMLs}*):List => (*\colorbox{lgray}{UMLs}*) [(*\colorbox{lyel}{P}*)-[#red]>(*\colorbox{lyel}{P}*):(*\colorbox{lblue}{A}*)] </n> 
      <s> (*\colorbox{lyel}{P}*) </s>
\end{lstlisting}
\end{lstcenter}
\caption{$\mathbb{K}$ snippet for HCSP to UML unidirectional transformation of rule 1 in Fig.~\ref{fig:example-frag}.}
\label{fig:uni-trans}
\end{figure}

\paragraph{Trustworthiness}
Beyond expressiveness, KBX establishes trust by proving the consistency of synchronized models for consumers. This is achieved using the {\K} framework, which generates Metamath proofs during synchronization to ensure models adhere to developer-defined {\K} specifications, checked by a matching logic proof checker (as elaborated in Section~\ref{sec: sim-veri-sync}). Leveraging Metamath's straightforward proof check mechanism, supported by multiple open-source implementations, this process facilitates easy and reliable certification of model consistency by both users and third-party entities.

\paragraph{Summary}
KBX offers an expressive framework for developers to construct formal synchronizers, coupled with a trustworthy method for consumers to verify model consistency. The following subsection will present a problem formulation for achieving simultaneous synchronization and verification. It will outline the technical challenges encountered and provide a roadmap for addressing these.

\subsection{Problem Statement}
\label{sec:problem-statement}

In this study, we tackle the {\Pid} ({\Pname}) problem, focusing on the verification and synchronization between model sets $M$ and $N$, exemplified by programs in HCSP and models in PlantUML. A key element to this problem is the complements $C$, indicative of information gaps inherent between $M$ and $N$ — crucial for differentiating BX from normal transformation.

{\centering
\begin{tcolorbox}[
    colback=lgray,
    colframe=darkgray,
    rounded corners,
    title={\Pid} Problem Statement,
    fonttitle=\bfseries\scshape,
    coltitle=white,
    boxrule=2pt,
    top=0mm, bottom=0mm, 
    width=\textwidth, height=10.5em, left=0mm, right=0mm,
    ]
Given a forward transformer $ux:M \rightarrow N$ compliant with formal definition $\Gamma^{ux}$, we aim to synthesize a formal synchronizer $\ell \in M \leftrightarrow N$ that comprises a forward transformer $putr:M \times C \rightarrow N \times C$, a backward transformer $putl:N \times C \rightarrow M \times C$, and a strategy for managing complements $C$. The transformers $putr$ and $putl$ should satisfy these conditions: (1) Compliance with their respective formal definitions $\Gamma^{putr}$ and $\Gamma^{putl}$, verified by a proof checker. (2) The output model \( n \in N \) from \( putr \) mirrors that of \( ux \) for identical input models, given an empty input complement. (3) Adherence to the round-tripping laws for reasonable synchronization.
\end{tcolorbox}
}

Following the problem statement, we distill the key to our work: (1) the formalization and verifiability of each definition for BX, and (2) the automatic provability of every executable tool derived from these definitions. We achieve this by synthesizing formal definitions within the {\K} framework, harnessing its capabilities for generating interpreters, verification tools and proofs. The input is a language-agnostic {\K} definition, covering syntax for both source and target transformations, state declarations, and transformation rules. This allows us to utilize existing {\K} semantics (e.g., Java, C, AADL) to validate related BX correctness and reuse portions of syntax and semantic rules, including static semantics, for BX construction. However, synthesizing formal BX definitions (i.e. forward and backward transformation definitions $\Gamma^{putr}$ and $\Gamma^{putl}$) and crafting a formal BX framework is far from trivial.

\textbf{Roadmap}.
Our work addresses the following key challenges in tackling the {\Pid} problem: 

\smallskip
\textit{Matching Logic-based BX Model} (Section~\ref{sec:model}). The {\K} framework, initially tailored for programming language semantics and verification, lacks direct support for BX formalization. Thus, our first step is to model the {\Pid} problem using matching logic to apply {\K}'s advanced formal methods, thereby satisfying the first {\Pid} condition. We strive to formalize a versatile model capable of handling multifaceted and multilingual BX scenarios beyond just HCSP and PlantUML, ensuring that our approach retains the expressive power of {\K} without imposing undue restrictions.

\smallskip
\textit{Forward Transformation Synthesis} (Section~\ref{sec:forward}). Building on our basis model, we seek to synthesize the forward transformer $putr$ addressing missing information recovery (e.g., recapturing the omitted PlantUML message color in HCSP). Altering the existing transformer $ux$ directly would contravene its formal definition $\Gamma^{ux}$, thus affecting its verifiability. We propose deriving a human-readable {\K} forward definition $\Gamma^{putr}$ from $\Gamma^{ux}$. This endeavor introduces two primary challenges: (1) Developing a method for manipulating complements $C$ within $\Gamma^{putr}$ that aids in information recovery while preserving {\K}'s language-agnostic nature to facilitate verified synchronization across various models. (2) Ensuring that the inclusion of complements manipulation in $putr$ maintains its output equivalence to $ux$ under identical inputs, satisfying the second {\Pid} condition.

\smallskip
\textit{Backward Transformation Synthesis} (Section~\ref{sec:backward}). Essential to our formal synchronizer, the backward transformer $putl$ confronts challenges in handling language-agnostic models and satisfying the round-tripping laws. We respond to these challenges by introducing the ``backward rewrite'' within matching logic, harnessing matching logic's capacity to transcend language specifics. This approach is crucial for synthesizing $\Gamma^{putl}$ and ensuring $putl$ satisfies the round-tripping laws.

\smallskip
\textit{Formal Synchronizer} (Section~\ref{sec: sim-veri-sync}). Although the formal transformers $putr$ and $putl$ are effectively derived from $\Gamma^{putr}$ and $\Gamma^{putl}$, the complexity of the {\K} framework and its toolchains presents challenges in practical application. To counter this, we introduce a formal synchronizer strategy that facilitates both synchronization and verification processes. This synchronizer not only streamlines the use of $putr$ and $putl$ as a symmetric lens but also generates consistency proofs for synchronized models (e.g., verifying if HCSP and UML models in Fig.~\ref{fig:example} represent the same system). These proofs, verifiable by a proof checker, offer tangible evidence of consistency, elevating the assurance level of the evaluation for safety-critical systems.

\smallskip
\textbf{Remark}.
Introducing BX to a formal framework presents several advantages: 

\textit{Trustworthy Synchronization with a Small Trust Base}. (1) \textit{Intrinsic Verifiability}. Definitions related to synchronization are readable, amenable and verifiable to customization and review. This diminishes the necessity for developing and trusting translations to other languages with a proof system. (2) \textit{Direct Correctness Verification}. This allows developers to define and verify critical transformation properties directly. Where operational semantics of synchronization targets are trustworthy, they can serve as an ideal basis for demonstrating synchronization correctness. (3) \textit{Trustworthy Execution}. The reliability of the synchronization process is ensured by generating proofs that are subsequently validated using a trusted proof checker.

\textit{Automatic Verification Comparable to Refinement Verification}. As discussed in Section~\ref{sec: sim-veri-sync}, we elucidate how verification through KBX aligns with, and is comparable to, refinement verification. This section also outlines methodologies for proving their compliance.

\section{Approach}
\label{sec:approach}

This section begins by formalizing the {\Pid} problem using matching logic (Section~\ref{sec:model}), laying the logical basis for the paper. We then introduce our solution, including three steps: the synthesis of forward (Section~\ref{sec:forward}) and backward (Section~\ref{sec:backward}) transformations, and their use in simultaneous verification and synchronization (Section~\ref{sec: sim-veri-sync}). Additionally, Section~\ref{sec: sim-veri-sync} explores the meaning of verification results in KBX.

\subsection{Matching Logic-based Bidirectional Transformation Model}
\label{sec:model}

This section presents the KBX model, the logical foundation of the KBX framework, to formulate the {\Pid} problem in $\mathbb{K}$ framework. This model includes two matching logic theories $\Gamma^{putr}$ and $\Gamma^{putl}$ by reforming the unidirectional one $\Gamma^{ux}$. Our synthesis uses this model to generate the formal BX definitions from the unidirectional ones in $\mathbb{K}$.

\smallskip
\textbf{KBX Input: Unidirectional Transformation Definition $\Gamma^{ux}$.} 
To provide an expressive and intuitive way to construct BX, the KBX model introduces a unidirectional definition $\Gamma^{ux}$ to capture the consistency definition within $\mathbb{K}$. This definition guides the transformations $ux : M \rightarrow N$ via matching logic rewriting $m\ \sigma_n\ \sigma_{s} \Rightarrow \sigma_m\ n\ \sigma_{s}$. These transformations begin with an input model $m$, an empty output model $\sigma_n$, and an initial transformation state $\sigma_{s}$. By substituting the rewrite rules in $\Gamma^{ux}$, these transformations result in an empty input model $\sigma_m$, an output model $n$, and a final transformation state $\sigma_{s}$. The following definition describes $\Gamma^{ux}$'s function and enables users to define consistency using $\mathbb{K}$ rewrite rules without limitations.

\begin{definition}\label{def:uni-trans}
Given the sorts $M$, $N$, $S$ and constants $\sigma_m\!:\!M, \sigma_n\!:\!N, \sigma_{s}\!:\!S$, unidirectional transformation definition $\Gamma^{ux}$ results in the following rewriting: 
    $$m\!:\!M, n\!:\!N.\ \Gamma^{ux} \vdash m\ \sigma_n\ \sigma_{s} \Rightarrow \sigma_m\ n\ \sigma_{s}$$
\end{definition}

In this context, the symbols $M$, $N$, and $S$ denote the syntax sorts for the input model, output model, and transformation helper, respectively. These sorts, such as the syntax of HCSP and plantUML, are defined within the $\Gamma^{ux}$ definition. Additionally, the definition includes a set of rewriting rules that facilitate the transformation process through the rewriting theory. The definition configuration declares the initial state and the structure of the states in the rewriting rules.

This formalization strikes a balance between specificity and generality. It is detailed enough to define a transformation, yet flexible enough to accommodate diverse languages and semantics. This is achieved without imposing limitations on the sorts $M, N$ (which define the syntax of transformation targets) and the rewrite rules in $\Gamma^{ux}$ (which determine the semantics of consistency). For instance, in Fig.~\ref{fig:uni-trans}, the syntax like \inlineK{HCSP} is captured by $M, N$. The \inlineK{configuration} (line 4) specifies the initial transformation state as $m\ \sigma_n\ \sigma_s$, and the \inlineK{rule} (lines 5-7) illustrates one of the $\Gamma^{ux}$ rules that govern the transformation process. Developers can adapt this model to other transformations using \inlineK{syntax} to define $M, N, S$, \inlineK{configuration} to set the transformation state, and \inlineK{rule} to delineate transformation semantics.

\smallskip
\textbf{KBX Output: Bidirectional Transformation Definitions $\Gamma^{putr}$ and $\Gamma^{putl}$.} We present Def.~\ref{def:mirror} and Def.~\ref{def:putrl-putlr} to describe the bidirectional transformation definitions in matching logic.

First, we clarify the concept of forward definition $\Gamma^{putr}$ by establishing its relation with $\Gamma^{ux}$. The following definition guarantees that when applying the transformations defined by $\Gamma^{ux}$ and $\Gamma^{putr}$ to an input model $m$, it will result in an identical transformed model $n$. Consequently, the forward definition $\Gamma^{putr}$ shares the same consistency definition as the unidirectional definition $\Gamma^{ux}$.

\begin{definition} \label{def:mirror}
Given the sorts $M$, $N$, $S$, $C$ and constants $\sigma_m\!:\!M, \sigma_n\!:\!N, \sigma_{s}\!:\!S$, the forward definition $\Gamma^{putr}$ mirrors the behavior of the unidirectional definition $\Gamma^{ux}$: $\forall m\!:\!M, n\!:\!N. \exists c\!:\!C, c^\prime\!:\!C. $
\begin{align*}
    \begin{tabular}{cc}
      $\Gamma^{ux} \vdash $ & $ m\ \sigma_n\ \sigma_{s} \Rightarrow \sigma_m\ n\ \sigma_{s}$  \\ \cline{1-2}
      $\Gamma^{putr} \vdash $ & $ m\ \sigma_n\ c\ \sigma_{s} \Rightarrow \sigma_m\ n\ c^\prime\ \sigma_{s}$ \\
    \end{tabular}
\end{align*}
\end{definition}

Second, we obtain the correct bidirectional definitions by refactoring the round-tripping laws from Definition~\ref{def:symmetric-lens} with matching logic results in Definition~\ref{def:putrl-putlr}. The definition $\Gamma^{putr}$ achieves the same directional transformation as $\Gamma^{ux}$, while $\Gamma^{putl}$ accomplishes the reverse transformation. Unlike the unidirectional definition $\Gamma^{ux}$, the bidirectional definitions $\Gamma^{putr}$ and $\Gamma^{putl}$ incorporate the sort $C$ and its corresponding configuration to retain information that might be lost during the transformation. Furthermore, according to \ref{def:putrl}, it is ensured that $\Gamma^{putr}$ accurately preserves information, and $\Gamma^{putl}$ effectively restores information. Similarly, \ref{def:putlr} guarantees that $\Gamma^{putl}$ correctly preserves information, and $\Gamma^{putr}$ aptly recovers information.

\begin{definition} \label{def:putrl-putlr}
Given the sorts $M$, $N$, $S$, $C$ and constants $\sigma_m\!:\!M, \sigma_n\!:\!N, \sigma_{s}\!:\!S$, bidirectional transformation definitions $\Gamma^{putr}$, $\Gamma^{putl}$ satisfy the following laws: $\forall m\!:\!M, n\!:\!N, c\!:\!C, c^\prime\!:\!C. $
  \begin{align*}
    \begin{tabular}{cc}
      $\Gamma^{putr} \vdash $ & $ m\ \sigma_n\ c\ \sigma_{s} \Rightarrow \sigma_m\ n\ c^\prime\ \sigma_{s}$  \\ \cline{1-2}
      $\Gamma^{putl} \vdash $ & $ \sigma_m\ n\ c^\prime\ \sigma_{s} \Rightarrow m\ \sigma_n\ c^\prime\ \sigma_{s}$ \\
    \end{tabular}\tag{P{\scriptsize ux}RL}\label{def:putrl} \\
    \begin{tabular}{cc}
      $\Gamma^{putl} \vdash $ & $ \sigma_m\ n\ c\ \sigma_{s} \Rightarrow m\ \sigma_n\ c^\prime\ \sigma_{s}$  \\ \cline{1-2}
      $\Gamma^{putr} \vdash $ & $ m\ \sigma_n\ c^\prime\ \sigma_{s} \Rightarrow \sigma_m\ n\ c^\prime\ \sigma_{s}$ \\
    \end{tabular}\tag{P{\scriptsize ux}LR}\label{def:putlr}
  \end{align*}
\end{definition}

In conclusion, we introduce the input definition $\Gamma^{ux}$ and the output definitions $\Gamma^{putr}, \Gamma^{putl}$ for KBX, aligning it with the scope of matching logic. Given that each matching logic definition corresponds to a $\mathbb{K}$ definition, developers can utilize the $\mathbb{K}$ framework to conveniently define consistency and implement the unidirectional transformation. This involves using the \textit{\textbf{syntax}} keyword to define sorts $M$, $N$, and $S$ through EBNF-like grammar; the \textit{\textbf{configuration}} keyword to identify constants $\sigma_m$, $\sigma_n$, $\sigma_{sr}$, and $\sigma_{sl}$ using XML-like grammar; the \textit{\textbf{rule}} keyword to establish rewrite/semantic rules within $\Gamma^{ux}$. By providing these $\mathbb{K}$ definitions, the $\mathbb{K}$ system automatically generates transformation and verification tools for automatic verification and synchronization.

Nevertheless, composing bidirectional transformation definitions $\Gamma^{putr}$, $\Gamma^{putl}$ poses challenges. Firstly, designing and implementing complements is time-consuming and error-prone. Secondly, managing numerous rewriting rules presents difficulties in ensuring accurate information storage and recovery. These factors contribute to a substantial workload and reduced reliability. Hence, we propose the subsequent synthesis to address these issues.
\subsection{Forward Transformation Synthesis}
\label{sec:forward}

This section delves into the design of a complements structure and the synthesis workflow for the forward transformation, denoted as $putr$, within the KBX framework.

\smallskip
\textbf{Structure of the Complements Holder}. Complements $C$ refers to the missing information to be recovered during the forward ($putr$) and backward ($putl$) transformation. In KBX, we manipulate this missing information in both directions of a BX by using a map that links common information to distinct information on either side of a rewrite rule, represented as \hForm{common \mapsto missing}. To illustrate this in the language-agnostic framework {\K}, consider the case of unidirectional transformation definition $\Gamma^{ux}$. This definition, as exemplified by lines 5-7 in Fig.~\ref{fig:uni-trans}, employs a set of rewrite rules to define transformation semantics. For each rule within {\K} definitions, various transformation-related information types are identifiable: (1) Variables/Tokens (e.g., \colorbox{lblue}{A}, \colorbox{lred}{L}, \inlineK{#black}), (2) Syntax productions like \inlineK{log(_)} and  \inlineK{_:=_} and , (3) State productions, which are syntactic sugars for generating \inlineK{Cell} patterns (e.g. \inlineK{<m> _ </m>}), (4) Side conditions that establish preconditions and postconditions, and (5) Rule attributes (e.g. \inlineK{priority(_)}, which determine the rule adoption order).

\smallskip
\textit{Discerning Missing Information}. We identify the \hForm{missing} information by examining the asymmetry around the rewrite symbol \inlineK{=>} within the {\K} definition $\Gamma$. This process involves dissecting $\Gamma$ into individual assessments of each rewrite rule. Such segmentation is reasonable because the transformation process comprises distinct \textit{snapshots}, each corresponding to the application of a rewrite rule (as detailed in Section~\ref{subsec:matching-logic}). Consequently, any information not present post-application of a rule is deemed $missing$ after the transformation. Our analysis across different information types leads to the following conclusions\footnote{Note that this analysis takes into account both directions of synchronization}:

\begin{itemize}
  \item Variables/Tokens are identified as $missing$ if they are not present on both sides of the rewrite symbol \inlineK{=>}. For instance, \colorbox{lred}{L} and \colorbox{lred}{R} in the $ux$ transformation targets are considered $missing$ as their values cannot be deduced from the right side of the rewrite rule.
  \item Syntax productions are considered $missing$ when there is an overlap in matched pattern sets on the right of two rewrite rules. This overlap impedes distinguishing syntax productions on the left side from the other. We presuppose such overlaps do not occur on the left side of $\Gamma^{ux}$'s rewrite rules to maintain deterministic transformation over symbolic results.
  \item Other information types like state productions, rule attributes, and side conditions are not categorized as $missing$. State productions and rule attributes remain constant through the rewriting process, and while side conditions influence rule application, they do not modify the state. Thus, they do not contribute to the $missing$ information during the transformation.
\end{itemize}

\textit{Designing Complements Holder}. 
Identifying $missing$ information is a critical initial step, but it's insufficient for forming a practical structure for information recovery. To address this, we use a \inlineK{Map} in the {\K} framework, structured as \hForm{common \mapsto missing}, ideal for data manipulation.

The \hForm{common} element represents shared information across both sides of the rewrite rules, facilitating consistent handling of the corresponding $missing$ information. However, defining what constitutes $common$ is crucial. Overburdening the \inlineK{Map} with excessive data leads to impractical key sizes. Therefore, we selectively identify $common$ elements:

\begin{itemize}
  \item \textit{Variables/Tokens}. Included in $common$ when appearing on both sides across different states, exemplified by \colorbox{lblue}{A} and \colorbox{lyel}{P} in Fig.~\ref{fig:uni-trans}. Conversely, variables/tokens common within the same state are excluded. For instance, \colorbox{lgray}{HCSPs} and \colorbox{lgray}{UMLs}, which merely serve as the context of rewrite rules, are excluded. Including such context-specific variables in $common$ would bloat $common$ and reduce its functionality, especially when slight changes in context render the complements ineffective for information recovery.
  \item \textit{Syntax Productions}. Ideally, $common$ could consist of left-hand syntax productions minus the $missing$. Yet, this approach is computationally intensive. To simplify, each rule in $common$ is assigned a unique ID. This facilitates the deterministic application of backward rewrite rules, ensuring the recovery of missing syntax productions when common variables differ. However, this approach introduces a limitation: it can result in indistinguishable syntax productions on the left side in scenarios where missing syntax productions exist, and common variables are identical across different rewrite rules. Despite this, it is reasonable, as identical syntax productions (i.e., right-hand productions' interactions) should inherently possess identical semantics, consequently leading to identical corresponding left-hand syntax productions rather than multiple discrete instances (i.e., missing syntax productions).
    \item \textit{Other information types.} State productions and rule attributes, similar to variables common in the same state, are excluded in $common$. Side conditions, analogous to syntax productions, are efficiently managed using rule ID.
\end{itemize}

Consequently, the state structure is \hForm{common \mapsto missing}, where \hForm{common} is a combination of rule ID and common variables/tokens across different states, and \hForm{missing} is different variables and tokens. This structure's generality lies in its reliance on information from the language-agnostic {\K} framework rather than specific transformation definitions. The rationale for this structure is further explored in Section~\ref{sec:backward}, where we discuss backward rewrite and synthesis.

\begin{figure}[t]
  \centering
  \includegraphics[width=\linewidth]{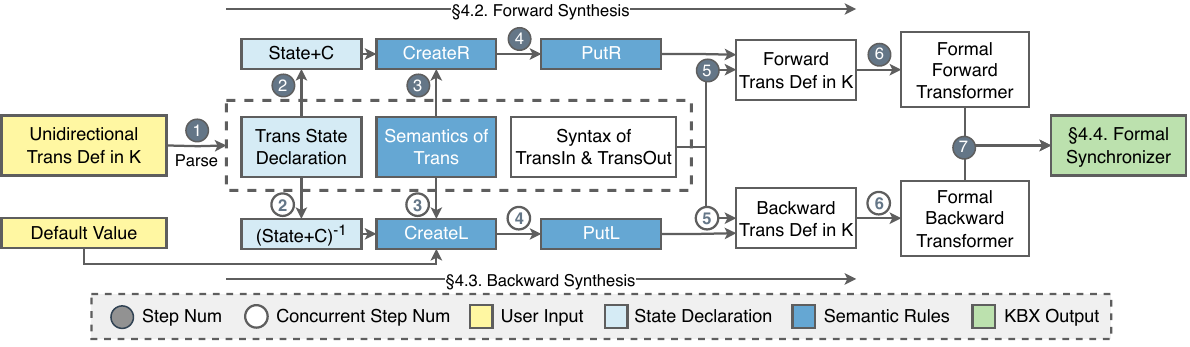}
  \caption{The Detailed Synthesis Workflow of KBX Formal BX Framework.}
  \Description[This workflow diagram demonstrates the process of parsing unidirectional transformation definitions, proceeding through multiple synthesis and transformation steps, and ultimately coordinating forward and backward transformations within a formalized programming language framework.]{This diagram illustrates a workflow for the definition and synchronization of transformations within a formal programming language environment. The process starts with the parsing of unidirectional transformation definitions in K language, followed by their extraction into an intermediate representation known as KAST. It then details the declaration of states and default values, the creation of left or right rules based on the semantics of transformations, and the placement of these rules within the syntactical structure of transformations. The workflow further defines forward or backward transformations in K, implements formal forward or backward transformations and culminates with a formal synchronizer ensuring consistency between them.}
  \label{fig:gen-workflow}
\end{figure}

\smallskip
\textbf{Workflow of Forward Transformation Synthesis}.
Figure~\ref{fig:gen-workflow} depicts the detailed synthesis workflow of KBX. We illustrate each step using Fig.~\ref{fig:forward-trans}, focusing on forward synthesis. The corresponding steps for backward synthesis are covered in Section~\ref{sec:backward}, and step 8 is elaborated in Section~\ref{sec: sim-veri-sync}. This example showcases a segment of the forward transformation definition $\Gamma^{putr}$, derived from the unidirectional $\Gamma^{ux}$ in Fig.~\ref{fig:uni-trans}. The \inlineK{configuration} keyword (line 1) sets up {\SaC}, and the \inlineK{rule} keyword (lines 4 to 7 and 9 to 12) defines rules in {\creater} and {\putr}, respectively. A key difference in Fig.~\ref{fig:forward-trans} is the \textit{complements manipulation} for recovering missing information.

\begin{figure}[ht]
\begin{lstcenter}[0.85\linewidth]
\begin{lstlisting}[language=K, escapeinside={(*}{*)}]
 // State+C
 configuration <m> $PGM:HCSP </m> <n> .K </n> <s> .K </s> <c> .Map </c>
 // CreateR: Complements Manipulation with Default Value
 rule <m> [log((*\colorbox{lblue}{A}*)), (*\colorbox{lred}{L}*) := (*\colorbox{lred}{R}*)] (*\colorbox{lgray}{HCSPs}*):List => (*\colorbox{lgray}{HCSPs}*) </m>
      <n> (*\colorbox{lgray}{UMLs}*):List => (*\colorbox{lgray}{UMLs}*) [(*\colorbox{lyel}{P}*)-[#black]>(*\colorbox{lyel}{P}*):(*\colorbox{lblue}{A}*)] </n> 
      <s> (*\colorbox{lyel}{P}*) </s>
      <c> Cp:Map =>  Cp [[1, (*\colorbox{lblue}{A}*), (*\colorbox{lyel}{P}*)] <- [[(*\colorbox{lred}{L}*), (*\colorbox{lred}{R}*)], [#black]]]</c>
 require Cp[[1, (*\colorbox{lblue}{A}*), (*\colorbox{lyel}{P}*)]] orDefault .List ==K .List
         orBool Cp[[1, (*\colorbox{lblue}{A}*), (*\colorbox{lyel}{P}*)]] orDefault .List ==K [[(*\colorbox{lred}{L}*), (*\colorbox{lred}{R}*)], [#black]]
 [priority(51)]
 // PutR: Complements Manipulation with Complements Holder
 rule <m> [log((*\colorbox{lblue}{A}*)), (*\colorbox{lred}{L}*) := (*\colorbox{lred}{R}*)] (*\colorbox{lgray}{HCSPs}*):List => (*\colorbox{lgray}{HCSPs}*) </m>
      <n> (*\colorbox{lgray}{UMLs}*):List => (*\colorbox{lgray}{UMLs}*) [(*\colorbox{lyel}{P}*)-[(*\colorbox{lpur}{C}*)]>(*\colorbox{lyel}{P}*):(*\colorbox{lblue}{A}*)] </n> 
      <s> (*\colorbox{lyel}{P}*) </s>
      <c> Cp:Map [1, (*\colorbox{lblue}{A}*), (*\colorbox{lyel}{P}*)] |-> [[ _ , _ ], [(*\colorbox{lpur}{C}*)]] 
       => Cp:Map [1, (*\colorbox{lblue}{A}*), (*\colorbox{lyel}{P}*)] |-> [[(*\colorbox{lred}{L}*), (*\colorbox{lred}{R}*)], [(*\colorbox{lpur}{C}*)]] </c>
\end{lstlisting}
\end{lstcenter}
\caption{The $\mathbb{K}$ definition for the forward transformation of rule 1.}
\label{fig:forward-trans}
\end{figure}

\textit{Step 1: $(syntax, state, rules) \gets Extract(Parse(\Gamma^{ux}))$}. The first step involves parsing the unidirectional transformation $\Gamma^{ux}$ into KAST, an intermediate representation in the {\K} framework. This parsing, aided by {\K}, ensures the transformation targets' syntax, like HCSP's \inlineK{_ ::= _}, is intuitively represented. Then, we extract critical components from the KAST: \hForm{syntax} (syntax definition, e.g., Fig.~\ref{fig:uni-trans} lines 1-3), \hForm{state} (transformation state declaration, e.g., Fig.~\ref{fig:uni-trans} line 4), and \hForm{rules} (rewrite/semantic rules, e.g., Fig.~\ref{fig:uni-trans} lines 5-7).

\textit{Step 2: $state^{+c} \gets AddCHolder(state)$}. This step adds a complements holder (\hForm{AddCHolder}) to the state for the forward transformation, differentiating it from the unidirectional state declaration. The complements holder, a $common \mapsto missing$ \inlineK{Map}, stores and retrieves missing information during the transformation. As shown in Fig.~\ref{fig:forward-trans} line 2, \hForm{AddCHolder} automatically generates the complements-holder cell with a distinct name \inlineK{c} and its initial state \inlineK{.Map}.

\textit{Step 3: $create^{r} \gets \forall \varphi\mbox{: }rule \in rules.\ Consist(\varphi\ CH(C => C(\varphi.common) \gets (\varphi.miss_r, \varphi.miss_l)))$}. This step involves enhancing each rewrite rule in the unidirectional transformation $\Gamma^{ux}$ with complements manipulation. The process, applied to all rules (\hForm{rules}), generates new rules (\hForm{create^{r}}) that mirror the original rules but include the storage of complements. Specifically, the \hForm{CH} function is employed to create a complements-holder pattern. This pattern dictates the rewriting of the complements holder's content (\hForm{C}) to capture both \hForm{\varphi.miss_r} (variables/tokens unique to the left-hand side) and \hForm{\varphi.miss_l} (unique to the right-hand side). If no $missing$ information (i.e. $\varphi.miss_r$, $\varphi.miss_l$) is detected, the original rule is retained unchanged. Additionally, the \hForm{Consist} function ensures no complements conflicts in the transformation source by verifying the state of $C(\varphi.common)$. To achieve this, we introduce a precondition to the rewrite rules in \hForm{create^{r}}. This precondition checks whether the set $C(\varphi.common)$ is either empty or if $C(\varphi.common)$ is equivalent to the $missing$ of the given rewrite rule, denoted as $C(\varphi.common) = () \lor C(\varphi.common) = (\varphi.miss_r, \varphi.miss_l)$.

Fig.~\ref{fig:forward-trans} illustrates our approach in lines 4-10, introducing complements manipulation to Fig.~\ref{fig:uni-trans} in lines 5-7. In line 7, a complements-holder state cell \inlineK{<c>} is utilized, driven by the \hForm{CH} function. This cell involves a rewrite pattern whose left-hand side is any possible complements holder \hForm{C} (i.e., \inlineK{Cp:Map}). The right-hand side is a map update operation (\inlineK{<-}). The map updates based on the key: \hForm{\varphi.common}, i.e., a list of rule ID (\inlineK{1}) and common variables/tokens (\colorbox{lblue}{A}, \colorbox{lyel}{P}). The updated value is a list of \hForm{\varphi.miss_r} (i.e., \colorbox{lred}{L} and \colorbox{lred}{R}) and \hForm{\varphi.miss_r} (i.e., \inlineK{#black}). The \hForm{Consist} function generates the side condition in lines 8-9. Line 10 introduces a lower priority for rule application, which is an implementation compromise for \hForm{HP} in the next step.

Note that, $C$ (e.g., \inlineK{Cp}) can be any holder without impacting the application of rule $\varphi$. Thus, the generated rewrite rules \hForm{create^{r}} guides equivalent transformation to the original \hForm{rules}.

\textit{Step 4: $put^{r} \gets \forall \varphi \in rules.\ let\ \varphi^\prime \gets T2V(\varphi)\ in\ HP(\varphi^\prime\ CH(C(\varphi.common) = (Any, \varphi^\prime.miss_l) =>\newline C(\varphi.common) = (\varphi^\prime.miss_r, \varphi^\prime.miss_l)))$}.
This step generates rewrite rules {\putr} (\hForm{put^{r}}), using the complements holder for information recovery. This recovery is achieved by the holder's left-hand side $\varphi^\prime.miss_l$. This rule also supports the modification of the synchronization source through $Any$ and only uses $\varphi.common$ as the identification for obtaining missing information. This synthesis step involves two new functions: \hForm{T2V} and \hForm{HP}. The \hForm{T2V} function converts all right-hand tokens into variables to use the complements holder for information recovery, rather than the default values. The \hForm{HP} function assigns these new rules a higher priority, ensuring they are applied before $create^{r}$ when matched complements exist.

Fig.~\ref{fig:forward-trans} demonstrates the generation of \hForm{put^{r}} in lines 12-16. This example mirrors Fig.~\ref{fig:uni-trans} (lines 5-7), except for the color token \inlineK{#black} is changed to the variable \colorbox{lpur}{C} using \hForm{T2V}. The complements-holder cell \inlineK{<c>} is then generated according to the new pattern \hForm{\varphi^\prime} using \hForm{CH}. As discussed before, the \hForm{HP} function provides a lower priority to the rules in $create^{r}$ to ensure the precedence of \hForm{put^{r}}. This compromise is because the default priority in {\K} is 50, which is also the highest priority.

Note that, the rules in \hForm{put^{r}} are not applicable when complements are empty. In such cases, the behavior of the transformation $putr$ follows that of $ux$ as per the rules in \hForm{create^{r}}.

\textit{Step 5: $\Gamma^{putr} \gets Print(syntax, state^{+c}, create^{r}, put^{r})$}. This step involves pretty printing the modified KAST into a readable {\K} definition for $\Gamma^{putr}$.

\textit{Step 6: $putr \gets Kompile(\Gamma^{putr})$}. The final step is compiling the $\Gamma^{putr}$ definition into a verifiable and executable forward transformer using \hForm{Kompile}.

In summary, the forward synthesis fulfills the following forward transformation requirements: (1) When complements are absent, the forward transformation $putr$ mirrors the unidirectional transformation $ux$ using rules in {\creater} (\hForm{create^{r}}). (2) With complements, $putr$ uses {\putr} (\hForm{putr^{r}}) to recover missing information. (3) All missing information $missing$ is retained in the complements holder \hForm{common \mapsto missing}, identifiable through $common$ between rewrite rules. (4) Transformation $putr$ validates the consistency of missing information in the transformation source using preconditions generated by \hForm{Consist}.  In essence, irrespective of whether the complements holder has retained the information, $putr$ maintains the same $missing$.

\subsection{Backward Transformation Synthesis}
\label{sec:backward}

Based on the complements holder and manipulation strategy, this section presents the synthesis basis and workflow for the backward transformation $putl$. 

\smallskip
\textbf{Theory of Backward Rewrite}:
The synthesis is grounded in the concept of \textit{backward rewrite}, denoted as $(_)^{-1} \in \Sigma$, aimed at formulating rewrite rules that adhere to the round-tripping laws in Definition~\ref{def:symmetric-lens}. Yet, the primary focus here is on the pure recovery aspect of the backward rewrite. The definition in matching logic reverses the left and right-hand sides of a rewrite rule $\varphi$, thus creating its backward counterpart.

\begin{definition}\label{def:backward-rewriting} 
  Given $\varphi \equiv \varphi_{lhs} \Rightarrow \varphi_{rhs}$, we define its backward rewrite rule, denoted as $(\varphi)^{-1}$ (abbreviated for the application $(\_)^{-1}\varphi$), as follows:
  $$
  (\varphi)^{-1} \equiv \varphi_{rhs} \Rightarrow \varphi_{lhs}
  $$
\end{definition}

This definition shows that the backward rewrite $(\_)^{-1}$ logically exchange $\varphi_{lhs}$ and $\varphi_{rhs}$ within the rewrite rule $\varphi$. Aggregating these backward rewrite rules forms the backward semantics $\Gamma^{-1}$, which aligns with the back-forth law (\ref{def:backforth}), enabling the recovery of initial states from final states in forward semantics $\Gamma$.

\begin{definition}
  Given any rewrite rule $\varphi_{lhs} \Rightarrow \varphi_{rhs} \in \Gamma$, we define the backward semantics as $(\Gamma)^{-1}$ as follows:
  $$ (\varphi_{lhs} \Rightarrow \varphi_{rhs})^{-1} \in (\Gamma)^{-1} $$
  This definition adheres to the back-forth law, which holds for all states $\varphi_{init}$ and $\varphi_{final}$:
  \begin{align*}
    \begin{tabular}{ll}
      $\hspace{0.4em}\Gamma\hspace{1.4em} \vdash$& $\varphi_{init}\hspace{0.6em} \Rightarrow \varphi_{final}$  \\ \cline{1-2} \addlinespace[0.3ex]
      $(\Gamma)^{-1} \vdash$& $\varphi_{final} \Rightarrow \varphi_{init}$ \\ 
    \end{tabular}\tag{B{\scriptsize ACK}F{\scriptsize ORTH}}\label{def:backforth}
  \end{align*}
\end{definition}

The satisfaction of the \ref{def:backforth} law can be understood through a two-step process. Firstly, we break down the forward execution, denoted as $\Gamma \vdash \varphi_{init} \Rightarrow \varphi_{final}$, into discrete sequential steps from $\varphi_i$ to $\varphi_{i+1}$. At each step $\varphi_i$, we identify a corresponding rewrite rule $\varphi_{lhs} \Rightarrow \varphi_{rhs}$ within $\Gamma$ and utilize a substitution $\theta_i$ such that $\varphi_{lhs}\theta_i \equiv \varphi_i$, thereby progressing to the subsequent step $\varphi_{i+1}$. Secondly, we initiate the backward execution commencing with the final state $\varphi_{final}$. At each subsequent step $\varphi_{i+1}$, we identify a corresponding backward rewrite rule $(\varphi_{lhs} \Rightarrow \varphi_{rhs})^{-1}$ within the backward semantics $(\Gamma)^{-1}$ and employ a substitution $\theta_{i+1}$ such that $\varphi_{rhs}\theta_{i+1} \equiv \varphi_{i+1}$, effectively regressing to the previous step $\varphi_i$. The construction of the backward execution, denoted as $(\Gamma)^{-1} \vdash \varphi_{final} \Rightarrow \varphi_{init}$, takes place step by step. Specifically, it is not necessary to present the reversed version of \ref{def:backforth} like the round-tripping laws in Def.~\ref{def:symmetric-lens} since $((\Gamma)^{-1})^{-1} = \Gamma$. 

\smallskip
\textbf{Workflow of Backward Transformation Synthesis}.
Leveraging the logical recovery potential of the backward rewrite, we follow Figure~\ref{fig:gen-workflow}'s steps to synthesize $putl$. Steps 1-2 are similar to forward transformation synthesis (discussed in Section~\ref{sec:forward}) and focus on subsequent steps. To clarify, we provide the following segment from the $\mathbb{K}$ definition for the backward transformation.

\begin{figure}[ht]
\begin{lstcenter}[0.86\linewidth]
\begin{lstlisting}[language=K, escapeinside={(*}{*)}]
 // (State+C(*\textcolor{kcomments}{$)^{-1}$}*)
 configuration <m> .K </m> <n> $PGM:K </n> <s> .K </s> <c> .Map </c>
 // CreateL: Complements Manipulation with Default Value
 rule <m> (*\colorbox{lgray}{HCSPs}*) => [log((*\colorbox{lblue}{A}*)), (*\fcolorbox{black}{white}{?1?}*) := (*\fcolorbox{black}{white}{?2?}*)] (*\colorbox{lgray}{HCSPs}*) </m>
      <n> (*\colorbox{lgray}{UMLs}*) [(*\colorbox{lyel}{P}*)-[(*\colorbox{lpur}{C}*)]>(*\colorbox{lyel}{P}*):(*\colorbox{lblue}{A}*)] => (*\colorbox{lgray}{UMLs}*) </n> 
      <s> (*\colorbox{lyel}{P}*) </s>
      <c> Cp:Map =>  Cp [[1, (*\colorbox{lblue}{A}*), (*\colorbox{lyel}{P}*)] <- [[(*\fcolorbox{black}{white}{?1?}*), (*\fcolorbox{black}{white}{?2?}*)], [(*\colorbox{lpur}{C}*)]]]</c>
 require Cp[[1, (*\colorbox{lblue}{A}*), (*\colorbox{lyel}{P}*)]] orDefault .List ==K .List
         orBool Cp[[1, (*\colorbox{lblue}{A}*), (*\colorbox{lyel}{P}*)]] orDefault .List ==K [[(*\fcolorbox{black}{white}{?1?}*), (*\fcolorbox{black}{white}{?2?}*)], [(*\colorbox{lpur}{C}*)]]
 [priority(51)]
 // PutL: Complements Manipulation with Complements Holder
 rule <m> (*\colorbox{lgray}{HCSPs}*) => [log((*\colorbox{lblue}{A}*)), (*\colorbox{lred}{L}*) := (*\colorbox{lred}{R}*)] (*\colorbox{lgray}{HCSPs}*) </m>
      <n> (*\colorbox{lgray}{UMLs}*) [(*\colorbox{lyel}{P}*)-[(*\colorbox{lpur}{C}*)]>(*\colorbox{lyel}{P}*):(*\colorbox{lblue}{A}*)] => (*\colorbox{lgray}{UMLs}*) </n> 
      <s> (*\colorbox{lyel}{P}*) </s>
      <c> Cp:Map [1, (*\colorbox{lblue}{A}*), (*\colorbox{lyel}{P}*)] |-> [[(*\colorbox{lred}{L}*), (*\colorbox{lred}{R}*)], [ _ ]] 
       => Cp:Map [1, (*\colorbox{lblue}{A}*), (*\colorbox{lyel}{P}*)] |-> [[(*\colorbox{lred}{L}*), (*\colorbox{lred}{R}*)], [(*\colorbox{lpur}{C}*)]] </c> 
\end{lstlisting}
\end{lstcenter}
\caption{The $\mathbb{K}$ definition for the backward transformation of rule 1.}
\label{fig:backward-trans}
\end{figure}

\textit{Step 2: $(state^{+c})^{-1} \gets ReverseIO(state^{+c})$}.
In this step, we utilize the state declaration $state^{+c}$ from the forward synthesis and apply \hForm{ReverseIO} to interchange the input and output cells. This interchange modifies their initial states without altering the overall configuration structure for rewrite rules. For example, as demonstrated in Fig.~\ref{fig:backward-trans} (line 2), \hForm{ReverseIO} identifies the input cell (\inlineK{<m>}) by its content, labeled as ``\$PGM'', which is then denoted as the terminal symbol \inlineK{.K}. Concurrently, the output cell (\inlineK{<n>}) is assigned the content ``\$PGM'' of type \inlineK{K}, which denotes any type defined in {\K} including the syntax of the transformation target.

In line with the back-forth law (\ref{def:backforth}), the unidirectional transformation $ux$ should conclude with $(state^{+c})^{-1}$ to ensure accurate recovery. Therefore, developers must ensure that the transformation $ux$ concludes with its initial state $state$, except the input cell becomes the terminal symbol \inlineK{.K}, and the output cell is adaptable to any model with type \inlineK{K}. Step 6 offers a modifiable definition, affording developers the flexibility to make alterations to $(state^{+c})^{-1}$ and thereby eliminating constraints on the final state of $ux$.

\textit{Step 3: $create^{l} \gets \forall \varphi\mbox{: }rule \in rules.\ F^4(VT2M(\varphi^{-1}))$}. This step enhances each rewrite rule in $\Gamma^{ux}$, integrating backward rewriting and complements manipulation. Firstly, we apply the backward rewrite, \hForm{(\_)^{-1}}, to each rule $\varphi$, creating its backward. In Fig.~\ref{fig:backward-trans} (lines 4-6, 12-15), this is shown as a reversed rewrite pattern within cells \textcolor{ksyntaxgreen}{<m>}, \textcolor{ksyntaxgreen}{<n>}, and \textcolor{ksyntaxgreen}{<s>}. This reversal also includes the contents of \textbf{require} and \textbf{ensure}, as well as ensuring consistency and standardization of variables within the rule. Secondly, the \hForm{VT2M} function transforms variables and tokens, appearing only on the right side of \inlineK{=>}, into question marks with unique identifiers. As demonstrated in Fig.~\ref{fig:backward-trans} line 4, variables \colorbox{lred}{L} and \colorbox{lred}{R} become \textcolor{black}{?1?} and \textcolor{black}{?2?}, respectively. This conversion enables users to define their {\defaultV} for unknown missing information. Lastly, \hForm{F^4} function mirrors the operation in Step 4 of forward synthesis, as shown in the following desugared format:

\begin{align*}
  create^{l} \gets & \forall \varphi\mbox{: }rule \in rules. \text{let}\ \varphi \gets VT2M(\varphi^{-1})\ \text{in} \\
  & Consist(\varphi\ CH(C => C(\varphi.common) \gets (\varphi.miss_r, \varphi.miss_l)))
\end{align*}

Note that, the interpretation of $\varphi.miss_r$ and $\varphi.miss_l$ is relative to the unidirectional transformation $ux$ direction. In the backward synthesis context, $\varphi.miss_r$ represents variables/tokens unique to the right-hand side, and $\varphi.miss_l$ to those unique to the left-hand side. Fig.~\ref{fig:backward-trans} lines 4-10 showcases the rule generated in this step.

\textit{Step 4: $put^l \gets ExAny((put^r)^{-1})$}.
In this step, we apply the backward rewrite to each rule in $put^r$, creating their backward counterparts in $put^l$. The \hForm{ExAny} function exchange the postion of $Any$ on the left-hand side of $put^r$, obtaining $(\varphi^\prime.miss_r, Any)$ for $put^l$. This process ensures complete adherence to the \ref{def:backforth}, enabling a seamless transition between states in the transformation trace for recovery. The application of this step is exemplified in Fig.~\ref{fig:backward-trans} (lines 12-15).

\textit{Step 5: $\Gamma^{putl} \gets Print(syntax, (state^{+c})^{-1}, create^{l}, put^{l})$}.
This step prints the modified KAST into a {\K} definition for $\Gamma^{putl}$.

\textit{Step 6: $putl \gets Kompile(\Gamma^{putl})$}.
Finally, $\Gamma^{putl}$ is compiled into an executable backward transformer $putl$ using $Kompile$.

In summary, these steps yield $putl$, the backward transformation, which, in conjunction with $putr$, achieves formal bidirectional transformation for simultaneous verification and synchronization. The round-tripping laws' satisfaction will be discussed in the following section.

\subsection{Formal Synchronizer for Simultaneous Synchronization and Verification}
\label{sec: sim-veri-sync}

\begin{figure}[t]
  \centering
  \includegraphics[width=0.8\linewidth]{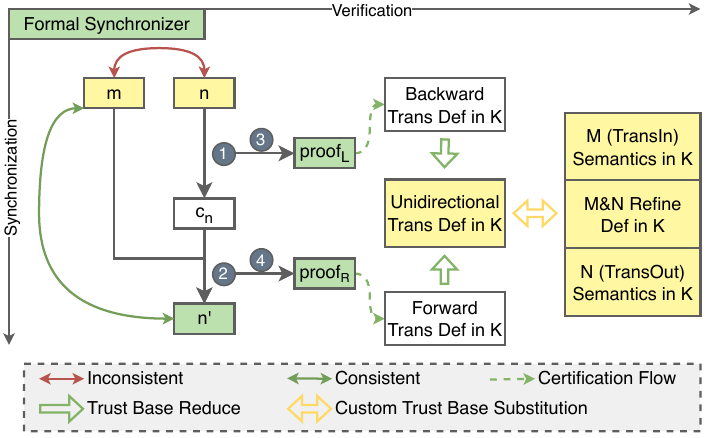}
\begin{tablenotes}
  \footnotesize
  \item[] Let $m$ and $n$ respectively represent an HCSP program and a UML model, with $c_n$ denoting missing UML information in HCSP (e.g., color). This information is extracted as per the {\K} definition $\Gamma^{putl}$, verified by $proof_L$. A synchronized UML model, $n^\prime$, is achieved, whose consistency with $m$ is confirmed by $proof_R$. Consistency between HCSP ($M$) and UML ($N$) is defined by $\Gamma^{ux}$, inherited by definitions $\Gamma^{putr}$ and $\Gamma^{putl}$.
\end{tablenotes}
  \caption{The Workflow of Formal Synchronizer.}
  \label{fig:use-workflow}
\end{figure}

This section introduces a formal synchronizer in the KBX framework, leveraging transformers $putr$ and $putl$. This synchronizer ensures simultaneous synchronization and verification of consistency relation $R$ between models $m$ and $n$, backed by formal proofs. It also aligns with the round-tripping laws (Definition~\ref{def:symmetric-lens}) to guarantee reasonable synchronization and establishes an equivalence between $R$ and refinement relation $R_{ref}$ through forward simulation. The workflow, including both synchronization and verification processes, is depicted in Fig.~\ref{fig:use-workflow}.

\smallskip
\textbf{Reasonable Synchronization Satisfying Round-tripping Laws}. The synchronizer inputs two models $m$ and $n$ and outputs synchronized models $m^\prime$ and $n^\prime$. Fig.~\ref{fig:use-workflow} illustrates a typical synchronization scenario where $m$ remains unchanged ($m = m^\prime$), and $n$ is adjusted to align with $m$. The synchronizer is versatile, allowing synchronization in both directions due to the commutative nature of $putr$ and $putl$. 
Step 1 involves applying $putl$ to model $n$, extracting missing information $c_n$ relative to model $m$. This step primarily uses rules from $create^l$ as no information in the complements holder matches rules in $put^l$. Step 2 employs $putr$, incorporating $c_n$ into the complements holder, to produce the consistent model $n^\prime$. This step relies on rules in $put^r$, where missing information is available in the complements holder.

These two steps are integrated into \hForm{putr} as outlined in Definition~\ref{def:symmetric-lens}, and reversing their order results in \hForm{putl}, ensuring compliance with the round-tripping laws. To elaborate, upon completing these steps, the output is \hForm{(n,c^{\prime})} as per \ref{def:oputlr}. Applying the transformer $putl$ to this output yields the same \hForm{(m,c^{\prime})}. The complements holder remains unchanged since steps 1 and 2 have already captured the information from both $m$ and $n$, and the rules in both $create^r$ and $put^r$ preserve identical information for the same source. Consequently, the original $m$ is retrieved using only the rules in $put^l$. The rules in $\Gamma^{putr}$ and $\Gamma^{putl}$ adhere to the \ref{def:backforth} law, barring the complements holder. For instance, regardless of whether a rule is in $create^{r}$ or $put^r$, applying $put^l$ retrieves the same left-hand side, leading back to $m$ from \hForm{(n,c^{\prime})}. Since the \ref{def:backforth} law is directionally neutral, this process is reversible, ensuring the round-tripping laws are fully met.

\smallskip
\textbf{Reliable Verification Comparable to Refinement Verification}. Steps 3-4 in the verification process verify the consistency between $m$ and $n^\prime$. Step 3 generates proof $proof_L$, confirming that the generation of $c_n$ aligns with $\Gamma^{putl}$ rules. Step 4 produces proof $proof_R$, verifying the consistency of $n^\prime$ with $m$ as per the {\K} definition $\Gamma^{putr}$. Both proof generation processes follow the methodologies presented in \cite{chenTrustworthySemanticsBasedLanguage2021}. These processes involve initially producing proof hints during transformation. Subsequently, we leverage these hints to construct Metamath proofs. These proofs encompass axioms derived from the {\K} definition, a proof goal that validates the correctness of the rewriting from $\varphi_{init}$ to $\varphi_{final}$, and a demonstration using the matching logic proof system and axioms to substantiate this goal. We then employ a Metamath checker to confirm that the generated proof satisfactorily resolves the proof goal. Proof generation, independent of transformer application order, is grounded in $\Gamma^{ux}$ and synthesis algorithms, forming the formal definition of consistency relation $R$.

\begin{definition} \label{def:R}
  The consistency relation $R$ between arbitrary models $m:M$ and $n:N$ is defined as follows,
  $$ (m,n) \in R \equiv \Gamma^{putr} \vdash m\ \sigma_n\ \sigma_c \ \sigma_{s} \Rightarrow \sigma_m\ n^\prime\ c\ \sigma_{s} \land  \Gamma^{putl} \vdash n\ \sigma_m\ c\ \sigma_{s} \Rightarrow \sigma_n\ m\ c^\prime\ \sigma_{s}$$
  or equivalently,
  $$(m,n) \in R \equiv \Gamma^{putl} \vdash n\ \sigma_m\ \sigma_c\ \sigma_{s} \Rightarrow \sigma_n\ m^\prime\ c\ \sigma_{s} \land \Gamma^{putr} \vdash m\ \sigma_n\ \ c \ \sigma_{s} \Rightarrow \sigma_m\ n\ c^\prime\ \sigma_{s}$$
          \end{definition}

The relation $R$ between models $m$ and $n$ has two equivalent definitions based on the application order of $\Gamma^{putr}$ and $\Gamma^{putl}$. In both of these definitions, the left side of the conjunction is employed for acquiring complements based on $\Gamma^{ux}$, while the right side of the conjunction incorporates the constancy defined within $\Gamma^{ux}$. To clarify the compliance, the refinement relation $R_{ref}$ is defined similarly but with operational semantics $\Gamma^M$ and $\Gamma^N$. 

\begin{definition}\label{def:refinement}
  The refinement relation $R_{ref}$ between arbitrary models $m:M$ and $n:N$ is defined as follows:
  \begin{align*}
         (m,n) \in R_{refine} &\equiv 
         (\Gamma^M \vdash m\ \sigma_{s_m} \Rightarrow \sigma_m\ s_m) \\
         &\land\ (\Gamma^N \vdash n\ \sigma_{s_n} \hspace{0.5em} \Rightarrow \sigma_n\ s_n) \\
         &\land\ R_S\ \sigma_{s_m}\ \sigma_{s_n} \land R_S\ s_m\ s_n 
  \end{align*}
  Here, $\Gamma^M$ and $\Gamma^N$ are the operational semantics for models $m$ and $n$, while $s_m$ and $s_n$ represent the execution states for these semantics. $\sigma_{s_m}$ and $\sigma_{s_n}$ refer to the initial states of $m$ and $n$, respectively, and $R_S$ defines the state relation predicate pattern. The semantics $\Gamma^M$ and $\Gamma^N$ are defined as follows: $\forall m\!:\!M, n\!:\!N.\ \exists s_m\!:\!S_M, s_n\!:\!S_N.$
  \begin{align*}
        &\Gamma^M \vdash m\ \sigma_{s_m} \Rightarrow \sigma_m\ s_m \
        &\Gamma^N \vdash n\ \sigma_{s_n} \Rightarrow \sigma_n\ s_n
  \end{align*}
\end{definition}

Hence, there are three types of compliance relations between consistency relation $R$ and refinement relation $R_{ref}$: (1) $R$ is an equivalent of $R_{ref}$ (written $R = R_{ref}$); (2) $R$ is a refinement of $R_{ref}$ (written $R \subset R_{ref}$); (3) $R_{ref}$ is a refinement of $R$ (written $R_{ref} \subset R$). Table~\ref{tab:verification-comparison} compares KBX-based verification and refinement verification for models $m$ and $n$.

\begin{table}[ht]
  \centering
  \caption{Comparison of KBX-based Verification and Refinement Verification.}
  \label{tab:verification-comparison}
  \begin{tabular}{ccc} 
  \toprule
   & \cellcolor[HTML]{EDEDED}KBX & Refinement \\ \midrule[0.8pt]
  Trust Base & \cellcolor[HTML]{EDEDED}$\Gamma^{putr}$, $\Gamma^{putl}$ (or $\Gamma^{ux}$, Synthesis) & $\Gamma^{M}$, $\Gamma^{N}$, $R_S$ \\ \hline
  Applicable & \cellcolor[HTML]{EDEDED}Complex M,N; Simple M,N Relation & Simple M,N; Complex M,N Relation \\ \hline
  Method & \cellcolor[HTML]{EDEDED}Verified Synchronization & Symbolic Execution or Theorem Proof\\ \hline
  On Failure & \cellcolor[HTML]{EDEDED}Automatic Alignment & Manual Alignment\\ \hline
  \end{tabular}
\end{table}

In summary, KBX provides a formal synchronizer that facilitates reasonable synchronization and reliable consistency verification between models $m$ and $n$.

\section{Evaluation}\label{sec:evaluation}

We have implemented the proposed approach in the KBX tool, which is built on top of the $\mathbb{K}$ formal framework. To assess the effectiveness of our KBX, we address the following research questions:

\begin{itemize}
\item \textbf{RQ1}: How does the development efficiency and trustworthiness of KBX compare to the existing BX frameworks?
\item \textbf{RQ2}: How does KBX perform in terms of both its expressiveness and trustworthiness as a verification approach?
\item \textbf{RQ3}: How can KBX facilitate the development of formal BX between UML and concurrent HCSP to address our industrial scenarios?
\end{itemize} \subsection{RQ1: Comparison with Existing BX Framework}
\label{sec:rq1}

In this section, we compare KBX against other BX frameworks using the criteria of development efficiency(E) and trustworthiness(T) in Table \ref{tab:satisfaction}. Concerning \cite{anjorinBenchmarkingBidirectionalTransformations2020,buchmannBXtendDSLLayeredFramework2022}, we choose typical and relevant BX frameworks for comparison, including BiGUL~\cite{koBiGULFormallyVerified2016,bettiniImplementingDomainspecificLanguages2016}, BiYacc~\cite{zhuBiYaccRollYour2015}, JTL~\cite{cicchettiJTLBidirectionalChange2011}, eMoflon~\cite{weidmannIncrementalBidirectionalModel2019b}, NMF~\cite{hinkelChangePropagationBidirectionality2019}, EVL+Strace~\cite{samimi-dehkordiEVLStraceNovel2018}, BXtend~\cite{buchmannBXtendAFrameworkBidirectional2018}+BXtendDSL~\cite{buchmannBXtendDSLLayeredFramework2022}, and Hobit~\cite{matsudaHobitProgrammingLenses2018}+Synbit~\cite{yamaguchiSynbitSynthesizingBidirectional2021}.
Specifically, we present the following assessment questions to qualify for the industrial requirements:

\begin{table}[t]
  \centering
  \caption{The satisfaction of BX frameworks in terms of \textbf{E}xpressiveness (E) and \textbf{T}rustworthiness (T).}
  \label{tab:satisfaction}
  \begin{threeparttable}
  \begin{tabularx}{\textwidth}{cCCCCCCCCCC}
      \toprule
        & KBX & \cellcolor[HTML]{EDEDED}BiGUL & BiYacc & \cellcolor[HTML]{EDEDED}JTL & eMoflon & \cellcolor[HTML]{EDEDED}NMF & EVL+ & \cellcolor[HTML]{EDEDED}BXtend & Hobit \\
      \midrule[1pt]
1 
& \cellcolor{lgreen}§\ref{sec:model}
& \cellcolor{lred}3
& \cellcolor{lred}6
& \cellcolor{lgreen}7
& \cellcolor{lgreen}3
& \cellcolor{lgreen}9
& \cellcolor{lgreen}3
& \cellcolor{lgreen}11
& \cellcolor{lred}12 \\
2 
& \cellcolor{lgreen}§\ref{sec:approach}
& \cellcolor{lgreen}4
& \cellcolor{lgreen}6
& \cellcolor{lgreen}7
& \cellcolor{lgreen}3
& \cellcolor{lgreen}9
& \cellcolor{lgreen}10
& \cellcolor{lgreen}11
& \cellcolor{lgreen}13 \\

3 
& \cellcolor{lgreen}1
& \cellcolor{lred}5
& \cellcolor{lgreen}6
& \cellcolor{lred}3
& \cellcolor{lred}3
& \cellcolor{lred}3
& \cellcolor{lred}3
& \cellcolor{lred}11
& \cellcolor{lred}12 \\

\hline
4 
& \cellcolor{lgreen}§\ref{sec:approach}
& \cellcolor{lred}5
& \cellcolor{lred}6
& \cellcolor{lred}8
& \cellcolor{lred}3
& \cellcolor{lred}3
& \cellcolor{lred}3
& \cellcolor{lred}11
& \cellcolor{lred}12 \\
5 
& \cellcolor{lgreen}1
& \cellcolor{lred}5
& \cellcolor{lred}6
& \cellcolor{lred}8
& \cellcolor{lred}3
& \cellcolor{lred}3
& \cellcolor{lred}3
& \cellcolor{lred}11
& \cellcolor{lred}12 \\
6 
& \cellcolor{lgreen}§\ref{sec:model}
& \cellcolor{lgreen}4
& \cellcolor{lgreen}6
& \cellcolor{lgreen}8
& \cellcolor{lgreen}3
& \cellcolor{lgreen}9
& \cellcolor{lred}3
& \cellcolor{lgreen}11
& \cellcolor{lgreen}12 \\
7 
& \cellcolor{lgreen}2
& \cellcolor{lred}5
& \cellcolor{lred}6
& \cellcolor{lred}8 & \cellcolor{lred}3 & \cellcolor{lred}3
& \cellcolor{lred}3
& \cellcolor{lred}3
& \cellcolor{lred}12\\
\bottomrule
\end{tabularx}
  \begin{tablenotes}
    \footnotesize
    \item [] 1 \cite{chenTrustworthySemanticsBasedLanguage2021}; 2 \cite{FrameworkTools2023}; 3 \cite{anjorinBenchmarkingBidirectionalTransformations2020}; 4 \cite{koBiGULFormallyVerified2016}; 5 \cite{huPrinciplesPracticeBidirectional2018}; 6 \cite{zhuUnifyingParsingReflective2020a};
    \item [] 7 \cite{cicchettiJTLBidirectionalChange2011}; 8 \cite{eramoEnhancingJTLTool2018}; 9 \cite{hinkelChangePropagationBidirectionality2019}; 10 \cite{samimi-dehkordiEVLStraceNovel2018};
    \item [] 11 \cite{buchmannBXtendDSLLayeredFramework2022}; 12 \cite{matsudaHobitProgrammingLenses2018}; 13 \cite{yamaguchiSynbitSynthesizingBidirectional2021} 
  \end{tablenotes}
\end{threeparttable}
\end{table}

\begin{itemize}
  \item[\textbf{E:}] The capability of constructing BX efficiently:
                \begin{itemize}
          \item[\textbf{1:}] Can we construct symmetric BX without implementing two asymmetric BX programs?
          \item \textit{Rationale:} Missing information on both sides of synchronization necessitates two asymmetric lenses, raising implementation costs and increasing the likelihood of errors in synchronizer development and maintenance.
          \item[\textbf{2:}] Can we construct BX in consistency definitions (e.g., unidirectional transformation definitions) without implementing the BX programs?
          \item \textit{Rationale:} Although every programming language can implement BX, the consistency definition streamlines this by merging two-sided transformation and complement concepts into one representation.
          \item[\textbf{3:}] Can we construct BX with BNF-like grammar without implementing parsers, printers, etc?
          \item \textit{Rationale:} Given that our intended synchronization targets are models or programs in various languages, the BX framework should automatically generate synchronizers that can manage these targets efficiently without additional effort.
                            \end{itemize}
                \item[\textbf{T:}] The capability of trustworthiness for verification:
        \begin{itemize}
          \item[\textbf{4:}] Can we define formal correctness specification for bidirectional transformation directly?
          \item \textit{Rationale:} If the answer is affirmative, BX developers can formally verify their BX programs, eliminating the implementation and trust in translating into formal frameworks.
          \item[\textbf{5:}] Can we verify the formal specification during the synchronization?
          \item \textit{Rationale:} This guarantees rigorous adherence of the synchronized targets to the user-defined formal specifications.
          \item[\textbf{6:}] Can we guarantee the BX behavior by explicit laws?
          \item \textit{Rationale:} BX without an explicit rule lacks a clear definition of rational synchronization.
          \item[\textbf{7:}] Can we validate the execution of BX with a proof checker?
          \item \textit{Rationale:} This question ensures that the synchronized targets adhere to the formal definition of consistency, which is reliable and can be readily verified by a third party.
        \end{itemize}
\end{itemize}

Table \ref{tab:satisfaction} displays the BX frameworks' evaluation results, with each cell indicating a framework's response to an assessment question. A ``Yes'' response is green and a ``No'' response is red. Every response in the table refers to specific evidence, indicated by a corresponding number at the table's bottom. Under the same evaluation methodology, we present the relation between our evidence with the assessment question as follows.

\begin{enumerate}
  \item Section~\ref{sec:model} captures the symmetric lenses.
  \item Sections \ref{sec:forward} and \ref{sec:backward} introduce the bidirectional transformation synthesis algorithm.
  \item Section~\ref{sec:model} takes $\mathbb{K}$ specifications as input and output, while $\mathbb{K}$ framework~\cite{FrameworkTools2023} can generate corresponding tools automatically. Additionally, \cite{chenTrustworthySemanticsBasedLanguage2021} presents a simple $\mathbb{K}$ definition with BNF-like grammar.
            \item Section~\ref{sec:model} shows that KBX's input is $\mathbb{K}$ definition that allows specifications for correctness like \cite{alpuenteAbstractContractSynthesis2020,hathhornDefiningUndefinedness2015}. Since the algorithms in Sections \ref{sec:forward} and \ref{sec:backward} focus on the complements recovery and don't modify the specification, the correctness specification also affects the bidirectional transformation.
  \item Section~\ref{sec:model} takes $\mathbb{K}$ specifications as input and output, while $\mathbb{K}$~\cite{FrameworkTools2023} allows symbolic execution for verification using these specifications.
  \item Section~\ref{sec:model} formalizes the round-tripping laws within matching logic. Sections~\ref{sec:forward} and~\ref{sec:backward}, present algorithms to synthesize bidirectional transformations that align with these laws.
  \item Sections~\ref{sec:forward} and~\ref{sec:backward} demonstrate KBX's ability to generate bidirectional transformation $\mathbb{K}$ specifications. Subsequently, we can employ these specifications to verify the execution of BX automatically using the method proposed in \cite{chenTrustworthySemanticsBasedLanguage2021}.
\end{enumerate}

To better illustrate KBX's efficacy in constructing BX, we employ the \textit{Family and Person} BX benchmark introduced by \cite{anjorinBenchmarkingBidirectionalTransformations2020}. We successfully passed all the benchmark tests, and the statistics for definition size are as follows:

\begin{table}[ht]
    \centering
    \caption{Size of the transformation definitions of all solutions.}
    \begin{tabularx}{\textwidth}{cCCCCCCCCCC}
      \toprule
        & KBX & \cellcolor[HTML]{EDEDED}BiGUL & BiYacc & \cellcolor[HTML]{EDEDED}JTL & eMoflon & \cellcolor[HTML]{EDEDED}NMF & EVL+ & \cellcolor[HTML]{EDEDED}BXtend & Hobit \\
      \midrule[0.8pt]
LOC 
& 35
& \cellcolor[HTML]{EDEDED}176
& -
& \cellcolor[HTML]{EDEDED}227
& 217
& \cellcolor[HTML]{EDEDED}279
& 1299
& \cellcolor[HTML]{EDEDED}211
& - \\
N words 
& 482
& \cellcolor[HTML]{EDEDED}1010
& -
& \cellcolor[HTML]{EDEDED}773
& 337
& \cellcolor[HTML]{EDEDED}607
& 2878
& \cellcolor[HTML]{EDEDED}565
& - \\
\bottomrule
    \end{tabularx}
    
  \begin{tablenotes}
    \footnotesize
    \item KBX synthesizes 2459 words definition for bidirectional transformation.
  \end{tablenotes}
    \label{tab:familiy-size}
\end{table}

The table presents several metrics, including Lines of Code (LOC) and word counts (N words), with a "-" indicating missing data. The data reveals that constructing BX with KBX requires \textit{the least human effort}. This benchmark exclusively employs JSON as both input and output formats and omits the line of code count for the parser and printer, which constitutes a significant portion. Indeed, the generated definition extends to 2,459 words, highlighting the substantial labor involved in implementing BX within the $\mathbb{K}$ framework without the assistance of KBX.

In summary, despite graphical limitations such as the absence of a graph language and interface, \textbf{KBX surpasses other BX frameworks for a specific set of queries}.

\subsection{RQ2: Capability as a Verification Approach}
\label{subsec:rq2}

In this section, we shift our attention from KBX synchronization capabilities to its verification potential. We will compare KBX with existing verification methods to answer two crucial questions: 
\begin{itemize}
    \item \textbf{RQ~2.1}:  What level of verification can we achieve through KBX expressiveness?
    \item \textbf{RQ~2.2}: How trustworthy are the outcomes of this verification process?
\end{itemize} 

\smallskip
\textbf{RQ~2.1}: We first evaluate the verification specification (i.e. input) of KBX. 
KBX employs the expressive language of the $\mathbb{K}$ framework for specification. This choice draws upon the proven expressiveness of the $\mathbb{K}$ framework as demonstrated in various language semantics, including Java~\cite{bogdanasKJavaCompleteSemantics2015} and x86-64~\cite{dasguptaCompleteFormalSemantics2019}. Hence, KBX empowers users with an intuitive and concise means of defining intricate transformations and valuable correctness specifications. An example of such a correctness specification is the handling of undefined behavior of C, as proposed by \cite{hathhornDefiningUndefinedness2015}.

Furthermore, beyond the need for additional correctness definitions, the unidirectional transformation definition itself can serve as a specification for consistency verification. This means that the synchronized models are verified consistent after bidirectional transformation. Additionally, the consistency relation aligns with the refinement relation, typically proven through forward simulation, a well-known approach in formal verification used by projects like seL4~\cite{kleinRefinementFormalVerification2010a}. More detailed information regarding this equivalence can be found in Section~\ref{sec: sim-veri-sync}. Consequently, the synchronized models also pass the refinement verification.

However, KBX introduces certain limitations on specification descriptions due to its rudimentary implementation. To ensure correct bidirectional transformation, the initial state must be equivalent to the termination state in the user-defined unidirectional transformation, except for the input and output cells. Furthermore, the implementation of backward rewrite lacks specialized handling of functions, necessitating $\mathbb{K}$ to have the capability to match the left-hand side of the rewriting that involves functions --- a feature currently not supported by $\mathbb{K}$.

\smallskip
\textbf{RQ~2.2}: To assess trustworthiness, we examine the trust base of KBX --- a critical factor necessary for establishing confidence in the verification outcome.

To begin with, we place our trust in the verification tool --- the $\mathbb{K}$ framework. \cite{chenTrustworthySemanticsBasedLanguage2021} have introduced an approach to minimize the trust base associated with the $\mathbb{K}$ framework. They achieve this by reducing it to two key components: the metamath proof checker and the matching logic theory within metamath. Metamath proof checkers consist of just a few hundred lines of code, yet they can efficiently verify thousands of theorems in mere seconds. Moreover, the matching logic theory comprises a concise 245 lines of code.

Another facet of trustworthiness pertains to the verification specifications, which serve as input for $\mathbb{K}$ and output of KBX. These specifications consist of the bidirectional transformation definitions written in $\mathbb{K}$, making them suitable for manual correctness validation. To enhance trustworthiness further, we can integrate handwritten unidirectional transformation and synthesis programs. Notably, unidirectional transformation definitions are more straightforward and concise compared to their bidirectional counterparts. The core implementation of the synthesis algorithm, after removing operations responsible for manipulating and extracting K syntax structures, contains fewer than 1000 lines of code.

In summary, \textbf{KBX has the capability to verify model correctness, consistency, and refinement with a minimal trust base and small constraints on the expressiveness of $\mathbb{K}$.}

\subsection{RQ3: Formal HCSP and UML BX for Industrial Scenarios}
\label{sec:rq3}

This section presents the first HCSP and UML BX, showcasing KBX's real-world effectiveness in addressing industrial challenges (Section \ref{sec:motivating-example}). This BX enables automatic synchronization of the HCSP and UML models and verification of their consistency. 
During our collaboration with an industry partner, we focus on modeling and validating a critical function in maglev train systems: partition handover. The function, encompassing 13 system components and 29 message types, requires seamless and secure transitions of maglev trains between rail segments and communication partitions. This necessity arises from the constraints of limited rail length and wireless communication range. Due to the use of different modeling languages, the development and verification of multiple models is tedious and error-prone. For example, communication and synchronization take up most of the time during the five iterations to construct the 15 UML diagrams and 10 HCSP scenarios. Furthermore, besides the 2 problems caused by understanding and 4 safety problems found by simulation, 
there are 58 severe problems owing to model inconsistency.

Therefore, to streamline our efforts and rigorously prove the consistency of this safety-critical system, we explore various methods for model synchronization and verification. Table~\ref{tab:comparison} outlines the costs before synchronization (Construction Costs), the expenses for maintaining and verifying consistency (Usage Costs), and the benefits resulting from synchronization and verification. Among these approaches, KBX offers an avenue for achieving verified synchronization with reduced costs while maintaining the trustworthiness of the formal method.

\begin{table}[h]
\centering
\caption{Comparison of Consistency Maintenance and Verification Approaches}
\label{tab:comparison}
{\small
\begin{tabular}{p{0.1\textwidth}p{0.3\textwidth}p{0.3\textwidth}p{0.2\textwidth}}
\toprule
\textbf{Approach} & \textbf{Construction Costs} & \textbf{Usage Costs} & \textbf{Benefits} \\
\midrule[0.8pt]
Manual & (1) Consistency definition documentation. & (1) Manual model synchronization. (2) Consistency documentation. & Traceable consistency of manual safeguards \\
\hline
Verified Translator & (1) translator from HCSP to UML. (2) translator from UML to HCSP. (3) verification of the translators. & (1) Automatic translation & Verified consistency without complements\\\hline
Forward simulation \& BX & (1) BX program for HCSP and UML synchronization. & (1) Automatic synchronization. (2) Verification and (3) Correction of the synchronized models. & Verified consistency \\\hline
Formal BX & 
(1) formal specifications of forward transformation and (2) backward transformation (3) verification of round-tripping laws & (1) Automatic synchronization with formal validation. & Verified consistency \\
\hline
KBX & 
(1) formal unidirectional transformation specification & (1) Automatic synchronization with formal validation. &  Verified consistency \\
\bottomrule
\end{tabular}}
\end{table}

Using KBX, we first construct the HCSP to UML $\mathbb{K}$ transformation, and then generate the UML and HCSP BX $\mathbb{K}$ transformation definitions automatically by the synthesizer based on Sections \ref{sec:forward} and \ref{sec:backward}. With this BX definition, $\mathbb{K}$ generates tools for synchronization and verification. 

\begin{table}[ht]
\centering
\caption{Support statistics for HCSP and plantUML.}
\begin{tabularx}{\textwidth}{c|cccccccc}
\toprule
\textbf{HCSP}
& skip
& assignment
& channel
& wait
& conditional
& interrupt
& hybrid   \\ \hline
\textbf{plantUML} 
& message
& comment
& loop
& option
& alt
& group
& color \\
\bottomrule
\end{tabularx}
\label{tab:hcsp-uml-support}
\end{table}

Table~\ref{tab:hcsp-uml-support} illustrates the features we have supported for HCSP and plantUML. The first row denotes the features supported for HCSP, while the second row pertains to the features supported for plantUML. While some features may not be included, the supported set consists of all crucial HCSP features, sufficient for modeling high-speed maglev train partition handover behavior. Next, we outline some of the technical steps involved in constructing the HCSP and UML BX: (1) Developing a unidirectional transformation from HCSP to plantUML. (2) Synthesizing a bidirectional transformation between HCSP and plantUML.

\smallskip
\textbf{Developing a unidirectional transformation from HCSP to plantUML}. First, we construct a transformation definition from HCSP to PlantUML within the $\mathbb{K}$ framework. We opt for this approach because it simplifies the process by starting with deterministic semantics while also offering ample support for HCSP features. This definition encompasses the syntax of both HCSP and PlantUML, defines the program configuration to describe the state structure and initial state, and formulates rewrite rules to specify the relation between HCSP and PlantUML.

\begin{figure}[t]
  \centering
  \includegraphics[width=0.7\textwidth]{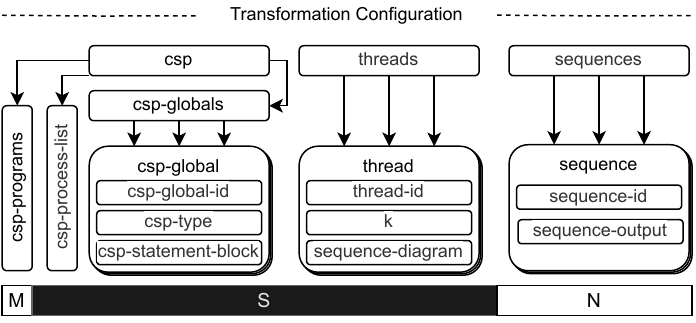}
  \caption{Transformation Configuration for UML and HCSP BX.}
  \vspace{-3mm}
  \label{fig:hcsp-uml-transformation-configuration}
\end{figure}

\smallskip
\textit{Program configuration}.
Fig.~\ref{fig:hcsp-uml-transformation-configuration} visually represents the configuration related to $M$, $N$, and $S$ discussed in Section~\ref{sec:model}. The HCSP programs to transform are contained in \textit{csp-programs}, while the transformed models in UML are stored in \textit{sequences}. The state $S$ is instantiated by \textit{csp-process-list}, \textit{csp-globals}, and \textit{threads}, initialized with empty ``$.$'' in sort ``$K$". 

\smallskip
\textit{Rewrite rules}. 
Constructing these rules presents challenges owing to semantic disparities. For instance, both HCSP and UML sequence diagrams incorporate communication, but HCSP applies channels for message transmission, while UML dispatches messages directly to objects. This divergence becomes even more intricate when addressing loops, interrupts, and other functionalities. Therefore, we introduce some of HCSP's concurrent semantics (i.e. spawn and join) to accurately describe the transformation process. The following phases offer a glimpse into our semantics.

\begin{enumerate}
  \item \textit{Initialization}: The first phase stores global information in \textit{csp-process-list} and \textit{csp-globals}. For each process or system, \textit{csp-globals} maintains the identifier, type (process or system), and statements, while \textit{csp-process-list} contains the original order of process identifiers as defined by the HCSP programs.
  \item \textit{Spawn}: The second phase spawns all the processes from \textit{csp-globals} and \textit{csp-process-list} to \textit{threads}.  Variables are assigned as follows: \textit{thread-id} is set to \textit{csp-global-id}, \textit{k} is set to \textit{csp-statement-block}, and \textit{sequence-diagram} is set to empty list $.List$.
  \item \textit{Transformation}: The third phase constructs the transformation from HCSP statements to UML ``statements'', which is the key part of the transformation specification.
  \item \textit{Join}: The forth phase join processes into \textit{sequences}. For each process joined, we output the corresponding \textit{sequence-diagram} to the \textit{sequence-output}.
  \item \textit{Termination}: The final phase standardizes UML in \textit{sequence-output} to obtain correct plantUML sequence diagram.
\end{enumerate}

\smallskip
\textbf{Synthesizing a bidirectional transformation between HCSP and plantUML}. Second, KBX derives bidirectional transformation definitions from the given unidirectional transformation definition.  Table~\ref{tab:generate-statistics} provides a comparison between transformations produced manually and those generated by KBX. Notably, KBX reduces code volume by 82.8\%, as calculated by the formula $\frac{TotalGenerated - Unidirectional}{TotalGenerated}$. Moreover, KBX simplifies the development process by handling complements and round-tripping laws, typically the most challenging aspects, relieving users' burdens.

\begin{table}[ht]
    \centering
    \caption{Size of the transformation definitions.}
    \begin{tabularx}{\textwidth}{cCCCC}
      \toprule
        & Unidirectional & Forward & Backward & Total Generated \\
      \midrule[0.8pt]
N words  
& 2105
& 4431
& 5790
& 10221 \\
\bottomrule
    \end{tabularx}
    \label{tab:generate-statistics}
\end{table}

In summary, \textbf{KBX is effective in the development of real-world formal HCSP and plantUML BX, significantly improving the efficiency of this process.}

\section{Related Work\label{sec:related}}

\ \ \ \ \textbf{Consistency Verification}. 
To verify complete systems, the utilization of multiple models is imperative due to limitations imposed by the language, the intricacy of the system itself, and the need to adhere to certain standards \cite{kulikSurveyPracticalFormal2022,reidMakingFormalMethods2020}. Researchers have employed various approaches, such as refinement verification, translation validation, and verified translators, to guarantee the consistency of these models.

Refinement verification~\cite{morganRefinementCalculus1988,backRefinementCalculusSystematic2012} is a common mechanism for ensuring consistency through forward simulation. For instance, \cite{kleinRefinementFormalVerification2010a,chenCompositionalVerificationInterruptible2018,liSecureFormallyVerified2021} verify the consistency of models across various abstract levels, ensuring the correctness of the operating system, device driver, and hypervisor.
Translation validation is another approach for consistency verification~\cite{lopesAlive2BoundedTranslation2021,bangSMTBasedTranslationValidation2022,sewellTranslationValidationVerified2013}, which automatically validates the correctness of the compilation process, ensuring the consistency between compilation sources and targets.
Previous studies~\cite{lanoFrameworkModelTransformation2015,leinenbachCompilerVerificationContext2008,leroyMechanizedSemanticsCompiler2012,srivastavaProgramVerificationProgram2010} proposed transformation verification methods to obtain verified translators. These approaches can automatically convert models while rigorously ensuring consistency. However, they only offer unidirectional transformations without the capability to recover missing information or perform reverse transformations.

\smallskip
\textbf{Bidirectional Transformation}. 
BX offers a practical solution for model synchronization and multiple BX frameworks have emerged~\cite{anjorinBenchmarkingBidirectionalTransformations2020,bettiniImplementingDomainspecificLanguages2016,buchmannBXtendAFrameworkBidirectional2018,buchmannBXtendDSLLayeredFramework2022,cicchettiJTLBidirectionalChange2011,koBiGULFormallyVerified2016,matsudaHobitProgrammingLenses2018,weidmannIncrementalBidirectionalModel2019b} that propose their bidirectional programming languages. For expressiveness, BiYacc~\cite{zhuBiYaccRollYour2015} presents a language-oriented language to roll the parser and reflective printer into one, but doubts arise about BiYacc's capability for BX between different languages due to the similarity between language syntax and its abstract syntax tree (AST). Hobit~\cite{matsudaHobitProgrammingLenses2018} enhances expressiveness by eliminating lens combinators, providing more flexibility in its representation. \citet{hinkelChangePropagationBidirectionality2019} took a different approach by reusing the C\# language to design a domain-specific language for bidirectional programming. For trustworthiness, \citet{koBiGULFormallyVerified2016} employ formal verification in BX to ensure the recovery of missing information. Going further, our work introduces formal bidirectional transformation to automatically maintain and verify model consistency from a formal unidirectional transformation specification. We reuse the language-oriented formal framework $\mathbb{K}$ without lens combinators to achieve both expressiveness and trustworthiness. While \cite{koBiGULFormallyVerified2016} also employs formal verification, our approach further verifies synchronized models by providing verified bidirectional transformers. This way, we reveal that the KBX is comparable with refinement verification for consistency. As discussed in Section~\ref{sec:rq1}, KBX is more expressive and trustworthy than existing BX frameworks.

\ToWrite{???Thus, different from \cite{koBiGULFormallyVerified2016} formalizing their bidirectional programming languages and verifying the rationale of }

To enhance the efficiency of developing BX programs, researchers have dedicated efforts to synthesizing BX through various approaches. For instance, \citet{yamaguchiSynbitSynthesizingBidirectional2021} use unidirectional sketches to synthesize bidirectional programs, while \citet{mainaSynthesizingQuotientLenses2018,miltnerSynthesizingBijectiveLenses2018,miltnerSynthesizingSymmetricLenses2019} employ a pair of regular expressions and example input-output pairs to achieve this automation. Given a partial specification, these algorithms are designed to narrow down the search space for identifying a suitable BX program within the context of existing BX frameworks. 
Compared to these recent works, KBX differs in that 
(1) it supports more flexible syntaxes and focuses on synthesizing formal BX specifications  (recall that the actual transformers are generated by $\mathbb{K}$);
(2) it uses formal unidirectional transformation specifications, without the need for additional input-output examples;
(3) it rests on our backward rewriting for deterministic and efficient synthesis, instead of previous type-directed enumerative algorithms.

\section{Conclusion\label{sec:conclusion}}

We have introduced KBX, the first formal bidirectional transformation framework for verified model synchronization in diverse languages and abstraction levels. Our evaluation shows that KBX outperforms other BX frameworks while maintaining the verification ability of $\mathbb{K}$ framework with verification outcomes comparable to refinement verification. Furthermore, we demonstrate the usability of KBX using a real-world application of HCSP \& UML BX.
For future work, we plan to apply KBX to more expressive languages and complicated specifications, thereby expanding its versatility and addressing more challenging scenarios.

\bibliography{reference.bib}


\begin{thebibliography}{59}


\ifx \showCODEN    \undefined \def \showCODEN     #1{\unskip}     \fi
\ifx \showDOI      \undefined \def \showDOI       #1{#1}\fi
\ifx \showISBNx    \undefined \def \showISBNx     #1{\unskip}     \fi
\ifx \showISBNxiii \undefined \def \showISBNxiii  #1{\unskip}     \fi
\ifx \showISSN     \undefined \def \showISSN      #1{\unskip}     \fi
\ifx \showLCCN     \undefined \def \showLCCN      #1{\unskip}     \fi
\ifx \shownote     \undefined \def \shownote      #1{#1}          \fi
\ifx \showarticletitle \undefined \def \showarticletitle #1{#1}   \fi
\ifx \showURL      \undefined \def \showURL       {\relax}        \fi
\providecommand\bibfield[2]{#2}
\providecommand\bibinfo[2]{#2}
\providecommand\natexlab[1]{#1}
\providecommand\showeprint[2][]{arXiv:#2}

\bibitem[\protect\citeauthoryear{??}{Com}{2017}]%
        {CommonCriteriaInformation2017b}
 \bibinfo{year}{2017}\natexlab{}.
\newblock \bibinfo{booktitle}{\emph{Common {{Criteria}} for {{Information
  Technology Security Evaluation}} - {{Part}} 3: {{Security}} Assurance
  Components} (\bibinfo{edition}{version 3.1 revision 5} ed.)}.
\newblock \bibinfo{publisher}{{CCMB-2017-04-003}},
  \bibinfo{address}{{https://www.commoncriteriaportal.org/cc/}}.
\newblock


\bibitem[\protect\citeauthoryear{??}{Pla}{2022}]%
        {PlantUML2022}
 \bibinfo{year}{2022}\natexlab{}.
\newblock \bibinfo{title}{{{PlantUML}}}.
\newblock \bibinfo{howpublished}{PlantUML}.
\newblock
\urldef\tempurl%
\url{https://github.com/plantuml/plantuml}
\showURL{%
\tempurl}


\bibitem[\protect\citeauthoryear{??}{Fra}{2023}]%
        {FrameworkTools2023}
 \bibinfo{year}{2023}\natexlab{}.
\newblock \bibinfo{title}{K {{Framework Tools}} 5.0}.
\newblock \bibinfo{howpublished}{Runtime Verification Inc.}.
\newblock
\urldef\tempurl%
\url{https://github.com/runtimeverification/k}
\showURL{%
\tempurl}


\bibitem[\protect\citeauthoryear{??}{Isa}{2024}]%
        {Isabelle}
 \bibinfo{year}{2024}\natexlab{}.
\newblock \bibinfo{title}{Isabelle}.
\newblock
\newblock
\urldef\tempurl%
\url{https://isabelle.in.tum.de/}
\showURL{%
\tempurl}


\bibitem[\protect\citeauthoryear{??}{Lea}{2024}]%
        {Lean}
 \bibinfo{year}{2024}\natexlab{}.
\newblock \bibinfo{title}{Lean}.
\newblock
\newblock
\urldef\tempurl%
\url{https://leanprover.github.io/}
\showURL{%
\tempurl}


\bibitem[\protect\citeauthoryear{??}{Wel}{2024}]%
        {WelcomeCoqProof}
 \bibinfo{year}{2024}\natexlab{}.
\newblock \bibinfo{title}{Welcome! | {{The Coq Proof Assistant}}}.
\newblock
\newblock
\urldef\tempurl%
\url{https://coq.inria.fr/}
\showURL{%
\tempurl}


\bibitem[\protect\citeauthoryear{Alpuente, Pardo, and Villanueva}{Alpuente
  et~al\mbox{.}}{2020}]%
        {alpuenteAbstractContractSynthesis2020}
\bibfield{author}{\bibinfo{person}{Mar{\'i}a Alpuente}, \bibinfo{person}{Daniel
  Pardo}, {and} \bibinfo{person}{Alicia Villanueva}.}
  \bibinfo{year}{2020}\natexlab{}.
\newblock \showarticletitle{Abstract {{Contract Synthesis}} and
  {{Verification}} in the {{Symbolic}} {$\mathbb{K}$} {{Framework}}}.
\newblock \bibinfo{journal}{\emph{Fundamenta Informaticae}}
  \bibinfo{volume}{177}, \bibinfo{number}{3-4} (\bibinfo{date}{Dec.}
  \bibinfo{year}{2020}), \bibinfo{pages}{235--273}.
\newblock
\showISSN{01692968, 18758681}
\urldef\tempurl%
\url{https://www.medra.org/servlet/aliasResolver?alias=iospress&doi=10.3233/FI-2020-1989}
\showURL{%
\tempurl}


\bibitem[\protect\citeauthoryear{Anjorin, Buchmann, Westfechtel, Diskin, Ko,
  Eramo, Hinkel, {Samimi-Dehkordi}, and Z{\"u}ndorf}{Anjorin
  et~al\mbox{.}}{2020}]%
        {anjorinBenchmarkingBidirectionalTransformations2020}
\bibfield{author}{\bibinfo{person}{Anthony Anjorin}, \bibinfo{person}{Thomas
  Buchmann}, \bibinfo{person}{Bernhard Westfechtel}, \bibinfo{person}{Zinovy
  Diskin}, \bibinfo{person}{Hsiang-Shang Ko}, \bibinfo{person}{Romina Eramo},
  \bibinfo{person}{Georg Hinkel}, \bibinfo{person}{Leila {Samimi-Dehkordi}},
  {and} \bibinfo{person}{Albert Z{\"u}ndorf}.} \bibinfo{year}{2020}\natexlab{}.
\newblock \showarticletitle{Benchmarking Bidirectional Transformations: Theory,
  Implementation, Application, and Assessment}.
\newblock \bibinfo{journal}{\emph{Software and systems modeling}}
  \bibinfo{volume}{19}, \bibinfo{number}{3} (\bibinfo{year}{2020}),
  \bibinfo{pages}{647--691}.
\newblock


\bibitem[\protect\citeauthoryear{Back and Wright}{Back and Wright}{2012}]%
        {backRefinementCalculusSystematic2012}
\bibfield{author}{\bibinfo{person}{Ralph-Johan Back} {and}
  \bibinfo{person}{Joakim Wright}.} \bibinfo{year}{2012}\natexlab{}.
\newblock \bibinfo{booktitle}{\emph{Refinement Calculus: A Systematic
  Introduction}}.
\newblock \bibinfo{publisher}{{Springer Science \& Business Media}}.
\newblock


\bibitem[\protect\citeauthoryear{Bang, Nam, Chun, Jhoo, and Lee}{Bang
  et~al\mbox{.}}{2022}]%
        {bangSMTBasedTranslationValidation2022}
\bibfield{author}{\bibinfo{person}{Seongwon Bang}, \bibinfo{person}{Seunghyeon
  Nam}, \bibinfo{person}{Inwhan Chun}, \bibinfo{person}{Ho~Young Jhoo}, {and}
  \bibinfo{person}{Juneyoung Lee}.} \bibinfo{year}{2022}\natexlab{}.
\newblock \showarticletitle{{{SMT-Based Translation Validation}} for {{Machine
  Learning Compiler}}}. In \bibinfo{booktitle}{\emph{Computer {{Aided
  Verification}}}}. \bibinfo{publisher}{{Springer, Cham}},
  \bibinfo{pages}{386--407}.
\newblock
\urldef\tempurl%
\url{https://linkspringer.53yu.com/chapter/10.1007/978-3-031-13188-2_19}
\showURL{%
\tempurl}


\bibitem[\protect\citeauthoryear{Bell}{Bell}{2006}]%
        {bellIntroductionIEC615082006}
\bibfield{author}{\bibinfo{person}{Ron Bell}.} \bibinfo{year}{2006}\natexlab{}.
\newblock \showarticletitle{Introduction to {{IEC}} 61508}. In
  \bibinfo{booktitle}{\emph{Acm International Conference Proceeding Series}},
  Vol.~\bibinfo{volume}{162}. \bibinfo{pages}{3--12}.
\newblock


\bibitem[\protect\citeauthoryear{Bettini and Efftinge}{Bettini and
  Efftinge}{2016}]%
        {bettiniImplementingDomainspecificLanguages2016}
\bibfield{author}{\bibinfo{person}{Lorenzo Bettini} {and} \bibinfo{person}{Sven
  Efftinge}.} \bibinfo{year}{2016}\natexlab{}.
\newblock \bibinfo{booktitle}{\emph{Implementing Domain-Specific Languages with
  {{Xtext}} and {{Xtend}}: Learn How to Implement a {{DSL}} with {{Xtext}} and
  {{Xtend}} Using Easy-to-Understand Examples and Best Practices}
  (\bibinfo{edition}{second edition} ed.)}.
\newblock \bibinfo{publisher}{{Packt Publishing}},
  \bibinfo{address}{{Birmingham Mumbai}}.
\newblock


\bibitem[\protect\citeauthoryear{Bogdanas and Ro{\c s}u}{Bogdanas and Ro{\c
  s}u}{2015}]%
        {bogdanasKJavaCompleteSemantics2015}
\bibfield{author}{\bibinfo{person}{Denis Bogdanas} {and}
  \bibinfo{person}{Grigore Ro{\c s}u}.} \bibinfo{year}{2015}\natexlab{}.
\newblock \showarticletitle{K-{{Java}}: {{A Complete Semantics}} of {{Java}}}.
  In \bibinfo{booktitle}{\emph{Proceedings of the 42nd {{Annual ACM
  SIGPLAN-SIGACT Symposium}} on {{Principles}} of {{Programming Languages}}}}.
  \bibinfo{publisher}{{ACM}}, \bibinfo{address}{{Mumbai India}},
  \bibinfo{pages}{445--456}.
\newblock
\urldef\tempurl%
\url{https://dl.acm.org/doi/10.1145/2676726.2676982}
\showURL{%
\tempurl}


\bibitem[\protect\citeauthoryear{Buchmann}{Buchmann}{2018}]%
        {buchmannBXtendAFrameworkBidirectional2018}
\bibfield{author}{\bibinfo{person}{Thomas Buchmann}.}
  \bibinfo{year}{2018}\natexlab{}.
\newblock \showarticletitle{{{BXtend-A Framework}} for ({{Bidirectional}})
  {{Incremental Model Transformations}}.}. In
  \bibinfo{booktitle}{\emph{{{MODELSWARD}}}}. \bibinfo{pages}{336--345}.
\newblock


\bibitem[\protect\citeauthoryear{Buchmann, Bank, and Westfechtel}{Buchmann
  et~al\mbox{.}}{2022}]%
        {buchmannBXtendDSLLayeredFramework2022}
\bibfield{author}{\bibinfo{person}{Thomas Buchmann}, \bibinfo{person}{Matthias
  Bank}, {and} \bibinfo{person}{Bernhard Westfechtel}.}
  \bibinfo{year}{2022}\natexlab{}.
\newblock \showarticletitle{{{BXtendDSL}}: {{A}} Layered Framework for
  Bidirectional Model Transformations Combining a Declarative and an Imperative
  Language}.
\newblock \bibinfo{journal}{\emph{Journal of Systems and Software}}
  \bibinfo{volume}{189} (\bibinfo{date}{July} \bibinfo{year}{2022}),
  \bibinfo{pages}{111288}.
\newblock
\showISSN{0164-1212}
\urldef\tempurl%
\url{https://www.sciencedirect.com/science/article/pii/S0164121222000462}
\showURL{%
\tempurl}


\bibitem[\protect\citeauthoryear{Chaochen, Ji, and Ravn}{Chaochen
  et~al\mbox{.}}{1996}]%
        {chaochenFormalDescriptionHybrid1996}
\bibfield{author}{\bibinfo{person}{Zhou Chaochen}, \bibinfo{person}{Wang Ji},
  {and} \bibinfo{person}{Anders~P. Ravn}.} \bibinfo{year}{1996}\natexlab{}.
\newblock \showarticletitle{A Formal Description of Hybrid Systems}. In
  \bibinfo{booktitle}{\emph{Hybrid {{Systems III}}: {{Verification}} and
  {{Control}} 3}}. \bibinfo{publisher}{{Springer}}, \bibinfo{pages}{511--530}.
\newblock


\bibitem[\protect\citeauthoryear{Chen, Wu, Shao, Lockerman, and Gu}{Chen
  et~al\mbox{.}}{2018}]%
        {chenCompositionalVerificationInterruptible2018}
\bibfield{author}{\bibinfo{person}{Hao Chen}, \bibinfo{person}{Xiongnan Wu},
  \bibinfo{person}{Zhong Shao}, \bibinfo{person}{Joshua Lockerman}, {and}
  \bibinfo{person}{Ronghui Gu}.} \bibinfo{year}{2018}\natexlab{}.
\newblock \showarticletitle{Toward Compositional Verification of Interruptible
  Os Kernels and Device Drivers}.
\newblock \bibinfo{journal}{\emph{Journal of Automated Reasoning}}
  \bibinfo{volume}{61}, \bibinfo{number}{1} (\bibinfo{year}{2018}),
  \bibinfo{pages}{141--189}.
\newblock


\bibitem[\protect\citeauthoryear{Chen, Lin, Trinh, and Ro{\c s}u}{Chen
  et~al\mbox{.}}{2021a}]%
        {chenTrustworthySemanticsBasedLanguage2021}
\bibfield{author}{\bibinfo{person}{Xiaohong Chen}, \bibinfo{person}{Zhengyao
  Lin}, \bibinfo{person}{Minh-Thai Trinh}, {and} \bibinfo{person}{Grigore Ro{\c
  s}u}.} \bibinfo{year}{2021}\natexlab{a}.
\newblock \showarticletitle{Towards a {{Trustworthy Semantics-Based Language
  Framework}} via {{Proof Generation}}}. In \bibinfo{booktitle}{\emph{Computer
  {{Aided Verification}}}} \emph{(\bibinfo{series}{Lecture {{Notes}} in
  {{Computer Science}}})}, \bibfield{editor}{\bibinfo{person}{Alexandra Silva}
  {and} \bibinfo{person}{K.~Rustan~M. Leino}} (Eds.).
  \bibinfo{publisher}{{Springer International Publishing}},
  \bibinfo{address}{{Cham}}, \bibinfo{pages}{477--499}.
\newblock


\bibitem[\protect\citeauthoryear{Chen, Lucanu, and Ro{\c s}u}{Chen
  et~al\mbox{.}}{2021b}]%
        {chenMatchingLogicExplained2021}
\bibfield{author}{\bibinfo{person}{Xiaohong Chen}, \bibinfo{person}{Dorel
  Lucanu}, {and} \bibinfo{person}{Grigore Ro{\c s}u}.}
  \bibinfo{year}{2021}\natexlab{b}.
\newblock \showarticletitle{Matching Logic Explained}.
\newblock \bibinfo{journal}{\emph{Journal of Logical and Algebraic Methods in
  Programming}}  \bibinfo{volume}{120} (\bibinfo{date}{April}
  \bibinfo{year}{2021}), \bibinfo{pages}{100638}.
\newblock
\showISSN{23522208}
\urldef\tempurl%
\url{https://linkinghub.elsevier.com/retrieve/pii/S2352220821000018}
\showURL{%
\tempurl}


\bibitem[\protect\citeauthoryear{Chen and Ro{\c s}u}{Chen and Ro{\c
  s}u}{2019}]%
        {chenMatchingMlogic2019}
\bibfield{author}{\bibinfo{person}{Xiaohong Chen} {and}
  \bibinfo{person}{Grigore Ro{\c s}u}.} \bibinfo{year}{2019}\natexlab{}.
\newblock \showarticletitle{Matching {$\mu$}-Logic}. In
  \bibinfo{booktitle}{\emph{2019 34th {{Annual ACM}}/{{IEEE Symposium}} on
  {{Logic}} in {{Computer Science}} ({{LICS}})}}. \bibinfo{publisher}{{IEEE}},
  \bibinfo{pages}{1--13}.
\newblock


\bibitem[\protect\citeauthoryear{Chen and Rosu}{Chen and Rosu}{2019}]%
        {chenMatchingMuLogicFoundation2019}
\bibfield{author}{\bibinfo{person}{Xiaohong Chen} {and}
  \bibinfo{person}{Grigore Rosu}.} \bibinfo{year}{2019}\natexlab{}.
\newblock \showarticletitle{Matching Mu-{{Logic}}: {{Foundation}} of {{K
  Framework}}}. In \bibinfo{booktitle}{\emph{8th {{Conference}} on {{Algebra}}
  and {{Coalgebra}} in {{Computer Science}} ({{CALCO}} 2019)}}.
  \bibinfo{publisher}{{Schloss Dagstuhl-Leibniz-Zentrum fuer Informatik}}.
\newblock


\bibitem[\protect\citeauthoryear{Cicchetti, Di~Ruscio, Eramo, and
  Pierantonio}{Cicchetti et~al\mbox{.}}{2011}]%
        {cicchettiJTLBidirectionalChange2011}
\bibfield{author}{\bibinfo{person}{Antonio Cicchetti}, \bibinfo{person}{Davide
  Di~Ruscio}, \bibinfo{person}{Romina Eramo}, {and} \bibinfo{person}{Alfonso
  Pierantonio}.} \bibinfo{year}{2011}\natexlab{}.
\newblock \showarticletitle{{{JTL}}: {{A Bidirectional}} and {{Change
  Propagating Transformation Language}}}. In \bibinfo{booktitle}{\emph{Software
  {{Language Engineering}}}} \emph{(\bibinfo{series}{Lecture {{Notes}} in
  {{Computer Science}}})}, \bibfield{editor}{\bibinfo{person}{Brian Malloy},
  \bibinfo{person}{Steffen Staab}, {and} \bibinfo{person}{Mark {van den
  Brand}}} (Eds.). \bibinfo{publisher}{{Springer}}, \bibinfo{address}{{Berlin,
  Heidelberg}}, \bibinfo{pages}{183--202}.
\newblock


\bibitem[\protect\citeauthoryear{Dasgupta, Park, Kasampalis, Adve, and Ro{\c
  s}u}{Dasgupta et~al\mbox{.}}{2019}]%
        {dasguptaCompleteFormalSemantics2019}
\bibfield{author}{\bibinfo{person}{Sandeep Dasgupta}, \bibinfo{person}{Daejun
  Park}, \bibinfo{person}{Theodoros Kasampalis}, \bibinfo{person}{Vikram~S.
  Adve}, {and} \bibinfo{person}{Grigore Ro{\c s}u}.}
  \bibinfo{year}{2019}\natexlab{}.
\newblock \showarticletitle{A Complete Formal Semantics of X86-64 User-Level
  Instruction Set Architecture}. In \bibinfo{booktitle}{\emph{Proceedings of
  the 40th {{ACM SIGPLAN Conference}} on {{Programming Language Design}} and
  {{Implementation}}}}. \bibinfo{publisher}{{ACM}}, \bibinfo{address}{{Phoenix
  AZ USA}}, \bibinfo{pages}{1133--1148}.
\newblock
\urldef\tempurl%
\url{https://dl.acm.org/doi/10.1145/3314221.3314601}
\showURL{%
\tempurl}


\bibitem[\protect\citeauthoryear{Ellison and {Grigore Ro{\c s}u}}{Ellison and
  {Grigore Ro{\c s}u}}{2011}]%
        {ellisonExecutableFormalSemantics2011}
\bibfield{author}{\bibinfo{person}{Chucky~M. Ellison} {and}
  \bibinfo{person}{{Grigore Ro{\c s}u}}.} \bibinfo{year}{2011}\natexlab{}.
\newblock \showarticletitle{An {{Executable Formal Semantics}} of {{C}} with
  {{Applications}}: {{Technical Report}}}.
\newblock \bibinfo{journal}{\emph{Acm Sigplan Notices}} (\bibinfo{year}{2011}).
\newblock
\urldef\tempurl%
\url{http://www.researchgate.net/publication/49175991_A_Formal_Semantics_of_C_with_Applications_Technical_Report}
\showURL{%
\tempurl}


\bibitem[\protect\citeauthoryear{Eramo, Pierantonio, and Tucci}{Eramo
  et~al\mbox{.}}{2018}]%
        {eramoEnhancingJTLTool2018}
\bibfield{author}{\bibinfo{person}{Romina Eramo}, \bibinfo{person}{Alfonso
  Pierantonio}, {and} \bibinfo{person}{Michele Tucci}.}
  \bibinfo{year}{2018}\natexlab{}.
\newblock \showarticletitle{Enhancing the {{JTL}} Tool for Bidirectional
  Transformations}. In \bibinfo{booktitle}{\emph{Conference {{Companion}} of
  the 2nd {{International Conference}} on {{Art}}, {{Science}}, and
  {{Engineering}} of {{Programming}}}}. \bibinfo{publisher}{{ACM}},
  \bibinfo{address}{{Nice France}}, \bibinfo{pages}{36--41}.
\newblock
\urldef\tempurl%
\url{https://dl.acm.org/doi/10.1145/3191697.3191720}
\showURL{%
\tempurl}


\bibitem[\protect\citeauthoryear{Hathhorn, Ellison, and Ro?u}{Hathhorn
  et~al\mbox{.}}{2015}]%
        {hathhornDefiningUndefinedness2015}
\bibfield{author}{\bibinfo{person}{Chris Hathhorn}, \bibinfo{person}{Chucky
  Ellison}, {and} \bibinfo{person}{Grigore Ro?u}.}
  \bibinfo{year}{2015}\natexlab{}.
\newblock \showarticletitle{Defining the Undefinedness of {{C}}}.
\newblock \bibinfo{journal}{\emph{ACM}} (\bibinfo{year}{2015}),
  \bibinfo{pages}{336--345}.
\newblock
\urldef\tempurl%
\url{http://dl.acm.org/doi/abs/10.1145/2737924.2737979}
\showURL{%
\tempurl}


\bibitem[\protect\citeauthoryear{Hinkel and Burger}{Hinkel and Burger}{2019}]%
        {hinkelChangePropagationBidirectionality2019}
\bibfield{author}{\bibinfo{person}{Georg Hinkel} {and} \bibinfo{person}{Erik
  Burger}.} \bibinfo{year}{2019}\natexlab{}.
\newblock \showarticletitle{Change Propagation and Bidirectionality in Internal
  Transformation {{DSLs}}}.
\newblock \bibinfo{journal}{\emph{Software \& Systems Modeling}}
  \bibinfo{volume}{18}, \bibinfo{number}{1} (\bibinfo{year}{2019}),
  \bibinfo{pages}{249--278}.
\newblock


\bibitem[\protect\citeauthoryear{Hofmann, Pierce, and Wagner}{Hofmann
  et~al\mbox{.}}{2011}]%
        {hofmannSymmetricLenses2011}
\bibfield{author}{\bibinfo{person}{Martin Hofmann}, \bibinfo{person}{Benjamin
  Pierce}, {and} \bibinfo{person}{Daniel Wagner}.}
  \bibinfo{year}{2011}\natexlab{}.
\newblock \showarticletitle{Symmetric Lenses}.
\newblock \bibinfo{journal}{\emph{ACM SIGPLAN Notices}} \bibinfo{volume}{46},
  \bibinfo{number}{1} (\bibinfo{year}{2011}), \bibinfo{pages}{371--384}.
\newblock


\bibitem[\protect\citeauthoryear{Hu and Ko}{Hu and Ko}{2018}]%
        {huPrinciplesPracticeBidirectional2018}
\bibfield{author}{\bibinfo{person}{Zhenjiang Hu} {and}
  \bibinfo{person}{Hsiang-Shang Ko}.} \bibinfo{year}{2018}\natexlab{}.
\newblock \showarticletitle{Principles and {{Practice}} of {{Bidirectional
  Programming}} in {{BiGUL}}}.
\newblock In \bibinfo{booktitle}{\emph{Bidirectional {{Transformations}}}},
  \bibfield{editor}{\bibinfo{person}{Jeremy Gibbons} {and}
  \bibinfo{person}{Perdita Stevens}} (Eds.). Vol.~\bibinfo{volume}{9715}.
  \bibinfo{publisher}{{Springer International Publishing}},
  \bibinfo{address}{{Cham}}, \bibinfo{pages}{100--150}.
\newblock
\urldef\tempurl%
\url{http://link.springer.com/10.1007/978-3-319-79108-1_4}
\showURL{%
\tempurl}


\bibitem[\protect\citeauthoryear{Jacklin}{Jacklin}{2012}]%
        {jacklinCertificationSafetyCriticalSoftware2012}
\bibfield{author}{\bibinfo{person}{Stephen Jacklin}.}
  \bibinfo{year}{2012}\natexlab{}.
\newblock \showarticletitle{Certification of {{Safety-Critical Software Under
  DO-178C}} and {{DO-278A}}}. In
  \bibinfo{booktitle}{\emph{Infotech@{{Aerospace}} 2012}}.
  \bibinfo{publisher}{{American Institute of Aeronautics and Astronautics}},
  \bibinfo{address}{{Garden Grove, California}}.
\newblock
\urldef\tempurl%
\url{https://arc.aiaa.org/doi/10.2514/6.2012-2473}
\showURL{%
\tempurl}


\bibitem[\protect\citeauthoryear{Klein, Andronick, Elphinstone, Murray, Sewell,
  Kolanski, and Heiser}{Klein et~al\mbox{.}}{2014}]%
        {kleinComprehensiveFormalVerification2014}
\bibfield{author}{\bibinfo{person}{Gerwin Klein}, \bibinfo{person}{June
  Andronick}, \bibinfo{person}{Kevin Elphinstone}, \bibinfo{person}{Toby
  Murray}, \bibinfo{person}{Thomas Sewell}, \bibinfo{person}{Rafal Kolanski},
  {and} \bibinfo{person}{Gernot Heiser}.} \bibinfo{year}{2014}\natexlab{}.
\newblock \showarticletitle{Comprehensive Formal Verification of an {{OS}}
  Microkernel}.
\newblock  (\bibinfo{year}{2014}).
\newblock
\urldef\tempurl%
\url{http://trustworthy.systems/publications/nictaabstracts/Klein_AEMSKH_14.abstract,
  /publications/nictaabstracts/Klein_AEMSKH_14.abstract}
\showURL{%
\tempurl}


\bibitem[\protect\citeauthoryear{Klein, Sewell, and Winwood}{Klein
  et~al\mbox{.}}{2010}]%
        {kleinRefinementFormalVerification2010a}
\bibfield{author}{\bibinfo{person}{Gerwin Klein}, \bibinfo{person}{Thomas
  Sewell}, {and} \bibinfo{person}{Simon Winwood}.}
  \bibinfo{year}{2010}\natexlab{}.
\newblock \showarticletitle{Refinement in the Formal Verification of the
  {{seL4}} Microkernel}.
\newblock In \bibinfo{booktitle}{\emph{Design and {{Verification}} of
  {{Microprocessor Systems}} for {{High-Assurance Applications}}}}.
  \bibinfo{publisher}{{Springer}}, \bibinfo{pages}{323--339}.
\newblock


\bibitem[\protect\citeauthoryear{Ko, Zan, and Hu}{Ko et~al\mbox{.}}{2016}]%
        {koBiGULFormallyVerified2016}
\bibfield{author}{\bibinfo{person}{Hsiang-Shang Ko}, \bibinfo{person}{Tao Zan},
  {and} \bibinfo{person}{Zhenjiang Hu}.} \bibinfo{year}{2016}\natexlab{}.
\newblock \showarticletitle{{{BiGUL}}: A Formally Verified Core Language for
  Putback-Based Bidirectional Programming}. In
  \bibinfo{booktitle}{\emph{Proceedings of the 2016 {{ACM SIGPLAN Workshop}} on
  {{Partial Evaluation}} and {{Program Manipulation}}}}.
  \bibinfo{publisher}{{ACM}}, \bibinfo{address}{{St. Petersburg FL USA}},
  \bibinfo{pages}{61--72}.
\newblock
\urldef\tempurl%
\url{https://dl.acm.org/doi/10.1145/2847538.2847544}
\showURL{%
\tempurl}


\bibitem[\protect\citeauthoryear{Kozen}{Kozen}{1983}]%
        {kozenResultsPropositionalMcalculus1983}
\bibfield{author}{\bibinfo{person}{Dexter Kozen}.}
  \bibinfo{year}{1983}\natexlab{}.
\newblock \showarticletitle{Results on the Propositional {$\mu$}-Calculus}.
\newblock \bibinfo{journal}{\emph{Theoretical computer science}}
  \bibinfo{volume}{27}, \bibinfo{number}{3} (\bibinfo{year}{1983}),
  \bibinfo{pages}{333--354}.
\newblock


\bibitem[\protect\citeauthoryear{Kulik, Dongol, Larsen, Macedo, Schneider,
  {Tran-J{\o}rgensen}, and Woodcock}{Kulik et~al\mbox{.}}{2022}]%
        {kulikSurveyPracticalFormal2022}
\bibfield{author}{\bibinfo{person}{Tomas Kulik}, \bibinfo{person}{Brijesh
  Dongol}, \bibinfo{person}{Peter~Gorm Larsen}, \bibinfo{person}{Hugo~Daniel
  Macedo}, \bibinfo{person}{Steve Schneider}, \bibinfo{person}{Peter~WV
  {Tran-J{\o}rgensen}}, {and} \bibinfo{person}{James Woodcock}.}
  \bibinfo{year}{2022}\natexlab{}.
\newblock \showarticletitle{A Survey of Practical Formal Methods for Security}.
\newblock \bibinfo{journal}{\emph{Formal Aspects of Computing}}
  \bibinfo{volume}{34}, \bibinfo{number}{1} (\bibinfo{year}{2022}),
  \bibinfo{pages}{1--39}.
\newblock


\bibitem[\protect\citeauthoryear{Lano, Clark, and {Kolahdouz-Rahimi}}{Lano
  et~al\mbox{.}}{2015}]%
        {lanoFrameworkModelTransformation2015}
\bibfield{author}{\bibinfo{person}{Kevin Lano}, \bibinfo{person}{Tony Clark},
  {and} \bibinfo{person}{S. {Kolahdouz-Rahimi}}.}
  \bibinfo{year}{2015}\natexlab{}.
\newblock \showarticletitle{A Framework for Model Transformation Verification}.
\newblock \bibinfo{journal}{\emph{Formal Aspects of Computing}}
  \bibinfo{volume}{27} (\bibinfo{year}{2015}), \bibinfo{pages}{193--235}.
\newblock


\bibitem[\protect\citeauthoryear{Leinenbach}{Leinenbach}{2008}]%
        {leinenbachCompilerVerificationContext2008}
\bibfield{author}{\bibinfo{person}{Dirk~Carsten Leinenbach}.}
  \bibinfo{year}{2008}\natexlab{}.
\newblock \showarticletitle{Compiler Verification in the Context of Pervasive
  System Verification}.
\newblock  (\bibinfo{year}{2008}).
\newblock


\bibitem[\protect\citeauthoryear{Leroy}{Leroy}{2012}]%
        {leroyMechanizedSemanticsCompiler2012}
\bibfield{author}{\bibinfo{person}{Xavier Leroy}.}
  \bibinfo{year}{2012}\natexlab{}.
\newblock \showarticletitle{Mechanized Semantics for Compiler Verification}. In
  \bibinfo{booktitle}{\emph{Programming {{Languages}} and {{Systems}}: 10th
  {{Asian Symposium}}, {{APLAS}} 2012, {{Kyoto}}, {{Japan}}, {{December}}
  11-13, 2012. {{Proceedings}} 10}}. \bibinfo{publisher}{{Springer}},
  \bibinfo{pages}{386--388}.
\newblock


\bibitem[\protect\citeauthoryear{Li, Li, Gu, Nieh, and Zhuang~Hui}{Li
  et~al\mbox{.}}{2021}]%
        {liSecureFormallyVerified2021}
\bibfield{author}{\bibinfo{person}{Shih-Wei Li}, \bibinfo{person}{Xupeng Li},
  \bibinfo{person}{Ronghui Gu}, \bibinfo{person}{Jason Nieh}, {and}
  \bibinfo{person}{John Zhuang~Hui}.} \bibinfo{year}{2021}\natexlab{}.
\newblock \showarticletitle{A {{Secure}} and {{Formally Verified Linux KVM
  Hypervisor}}}. In \bibinfo{booktitle}{\emph{2021 {{IEEE Symposium}} on
  {{Security}} and {{Privacy}} ({{SP}})}}. \bibinfo{publisher}{{IEEE}},
  \bibinfo{address}{{San Francisco, CA, USA}}, \bibinfo{pages}{1782--1799}.
\newblock
\urldef\tempurl%
\url{https://ieeexplore.ieee.org/document/9519433/}
\showURL{%
\tempurl}


\bibitem[\protect\citeauthoryear{Liu, Lv, Quan, Zhan, Zhao, Zhou, and Zou}{Liu
  et~al\mbox{.}}{2010}]%
        {liuCalculusHybridCSP2010}
\bibfield{author}{\bibinfo{person}{Jiang Liu}, \bibinfo{person}{Jidong Lv},
  \bibinfo{person}{Zhao Quan}, \bibinfo{person}{Naijun Zhan},
  \bibinfo{person}{Hengjun Zhao}, \bibinfo{person}{Chaochen Zhou}, {and}
  \bibinfo{person}{Liang Zou}.} \bibinfo{year}{2010}\natexlab{}.
\newblock \showarticletitle{A Calculus for Hybrid {{CSP}}}. In
  \bibinfo{booktitle}{\emph{Programming {{Languages}} and {{Systems}}: 8th
  {{Asian Symposium}}, {{APLAS}} 2010, {{Shanghai}}, {{China}}, {{November}}
  28-{{December}} 1, 2010. {{Proceedings}} 8}}.
  \bibinfo{publisher}{{Springer}}, \bibinfo{pages}{1--15}.
\newblock


\bibitem[\protect\citeauthoryear{Lopes, Lee, Hur, Liu, and Regehr}{Lopes
  et~al\mbox{.}}{2021}]%
        {lopesAlive2BoundedTranslation2021}
\bibfield{author}{\bibinfo{person}{Nuno~P. Lopes}, \bibinfo{person}{Juneyoung
  Lee}, \bibinfo{person}{Chung-Kil Hur}, \bibinfo{person}{Zhengyang Liu}, {and}
  \bibinfo{person}{John Regehr}.} \bibinfo{year}{2021}\natexlab{}.
\newblock \showarticletitle{Alive2: Bounded Translation Validation for
  {{LLVM}}}. In \bibinfo{booktitle}{\emph{Proceedings of the 42nd {{ACM SIGPLAN
  International Conference}} on {{Programming Language Design}} and
  {{Implementation}}}}. \bibinfo{publisher}{{ACM}}, \bibinfo{address}{{Virtual
  Canada}}, \bibinfo{pages}{65--79}.
\newblock
\urldef\tempurl%
\url{https://dl.acm.org/doi/10.1145/3453483.3454030}
\showURL{%
\tempurl}


\bibitem[\protect\citeauthoryear{Maina, Miltner, Fisher, Pierce, Walker, and
  Zdancewic}{Maina et~al\mbox{.}}{2018}]%
        {mainaSynthesizingQuotientLenses2018}
\bibfield{author}{\bibinfo{person}{Solomon Maina}, \bibinfo{person}{Anders
  Miltner}, \bibinfo{person}{Kathleen Fisher}, \bibinfo{person}{Benjamin~C.
  Pierce}, \bibinfo{person}{David Walker}, {and} \bibinfo{person}{Steve
  Zdancewic}.} \bibinfo{year}{2018}\natexlab{}.
\newblock \showarticletitle{Synthesizing Quotient Lenses}.
\newblock \bibinfo{journal}{\emph{Proceedings of the ACM on Programming
  Languages}} \bibinfo{volume}{2}, \bibinfo{number}{ICFP} (\bibinfo{date}{July}
  \bibinfo{year}{2018}), \bibinfo{pages}{1--29}.
\newblock
\showISSN{2475-1421}
\urldef\tempurl%
\url{https://dl.acm.org/doi/10.1145/3236775}
\showURL{%
\tempurl}


\bibitem[\protect\citeauthoryear{Matsuda and Wang}{Matsuda and Wang}{2018}]%
        {matsudaHobitProgrammingLenses2018}
\bibfield{author}{\bibinfo{person}{Kazutaka Matsuda} {and}
  \bibinfo{person}{Meng Wang}.} \bibinfo{year}{2018}\natexlab{}.
\newblock \showarticletitle{Hobit: {{Programming}} Lenses without Using Lens
  Combinators}. In \bibinfo{booktitle}{\emph{European {{Symposium}} on
  {{Programming}}}}. \bibinfo{publisher}{{Springer}}, \bibinfo{pages}{31--59}.
\newblock


\bibitem[\protect\citeauthoryear{Miltner, Fisher, Pierce, Walker, and
  Zdancewic}{Miltner et~al\mbox{.}}{2018}]%
        {miltnerSynthesizingBijectiveLenses2018}
\bibfield{author}{\bibinfo{person}{Anders Miltner}, \bibinfo{person}{Kathleen
  Fisher}, \bibinfo{person}{Benjamin~C. Pierce}, \bibinfo{person}{David
  Walker}, {and} \bibinfo{person}{Steve Zdancewic}.}
  \bibinfo{year}{2018}\natexlab{}.
\newblock \showarticletitle{Synthesizing Bijective Lenses}.
\newblock \bibinfo{journal}{\emph{Proceedings of the ACM on Programming
  Languages}} \bibinfo{volume}{2}, \bibinfo{number}{POPL} (\bibinfo{date}{Jan.}
  \bibinfo{year}{2018}), \bibinfo{pages}{1--30}.
\newblock
\showISSN{2475-1421}
\urldef\tempurl%
\url{https://dl.acm.org/doi/10.1145/3158089}
\showURL{%
\tempurl}


\bibitem[\protect\citeauthoryear{Miltner, Maina, Fisher, Pierce, Walker, and
  Zdancewic}{Miltner et~al\mbox{.}}{2019}]%
        {miltnerSynthesizingSymmetricLenses2019}
\bibfield{author}{\bibinfo{person}{Anders Miltner}, \bibinfo{person}{Solomon
  Maina}, \bibinfo{person}{Kathleen Fisher}, \bibinfo{person}{Benjamin~C.
  Pierce}, \bibinfo{person}{David Walker}, {and} \bibinfo{person}{Steve
  Zdancewic}.} \bibinfo{year}{2019}\natexlab{}.
\newblock \showarticletitle{Synthesizing Symmetric Lenses}.
\newblock \bibinfo{journal}{\emph{Proceedings of the ACM on Programming
  Languages}} \bibinfo{volume}{3}, \bibinfo{number}{ICFP} (\bibinfo{date}{July}
  \bibinfo{year}{2019}), \bibinfo{pages}{1--28}.
\newblock
\showISSN{2475-1421}
\urldef\tempurl%
\url{https://dl.acm.org/doi/10.1145/3341699}
\showURL{%
\tempurl}


\bibitem[\protect\citeauthoryear{Morgan, Robinson, and Gardiner}{Morgan
  et~al\mbox{.}}{1988}]%
        {morganRefinementCalculus1988}
\bibfield{author}{\bibinfo{person}{Carroll Morgan}, \bibinfo{person}{Ken
  Robinson}, {and} \bibinfo{person}{Paul Gardiner}.}
  \bibinfo{year}{1988}\natexlab{}.
\newblock \bibinfo{booktitle}{\emph{On the Refinement Calculus}}.
\newblock Number~70 in \bibinfo{series}{Technical Monograph / {{Oxford Univ}}.
  {{Computing Laboratory}}, {{Programming Research Group}}}.
  \bibinfo{publisher}{{University Computing Laboratory}},
  \bibinfo{address}{{Oxford}}.
\newblock


\bibitem[\protect\citeauthoryear{Park, Stef{\u a}nescu, and Ro{\c s}u}{Park
  et~al\mbox{.}}{2015}]%
        {parkKJSCompleteFormal2015}
\bibfield{author}{\bibinfo{person}{Daejun Park}, \bibinfo{person}{Andrei
  Stef{\u a}nescu}, {and} \bibinfo{person}{Grigore Ro{\c s}u}.}
  \bibinfo{year}{2015}\natexlab{}.
\newblock \showarticletitle{{{KJS}}: A Complete Formal Semantics of
  {{JavaScript}}}. In \bibinfo{booktitle}{\emph{Proceedings of the 36th {{ACM
  SIGPLAN Conference}} on {{Programming Language Design}} and
  {{Implementation}}}} \emph{(\bibinfo{series}{{{PLDI}} '15})}.
  \bibinfo{publisher}{{Association for Computing Machinery}},
  \bibinfo{address}{{New York, NY, USA}}, \bibinfo{pages}{346--356}.
\newblock
\urldef\tempurl%
\url{https://doi.org/10.1145/2737924.2737991}
\showURL{%
\tempurl}


\bibitem[\protect\citeauthoryear{Reid, Church, Flur, {de Haas}, Johnson, and
  Laurie}{Reid et~al\mbox{.}}{2020}]%
        {reidMakingFormalMethods2020}
\bibfield{author}{\bibinfo{person}{Alastair Reid}, \bibinfo{person}{Luke
  Church}, \bibinfo{person}{Shaked Flur}, \bibinfo{person}{Sarah {de Haas}},
  \bibinfo{person}{Maritza Johnson}, {and} \bibinfo{person}{Ben Laurie}.}
  \bibinfo{year}{2020}\natexlab{}.
\newblock \showarticletitle{Towards Making Formal Methods Normal: Meeting
  Developers Where They Are}.
\newblock \bibinfo{journal}{\emph{arXiv preprint arXiv:2010.16345}}
  (\bibinfo{year}{2020}).
\newblock
\showeprint[arxiv]{2010.16345}


\bibitem[\protect\citeauthoryear{{Samimi-Dehkordi}, Zamani, and
  {Kolahdouz-Rahimi}}{{Samimi-Dehkordi} et~al\mbox{.}}{2018}]%
        {samimi-dehkordiEVLStraceNovel2018}
\bibfield{author}{\bibinfo{person}{Leila {Samimi-Dehkordi}},
  \bibinfo{person}{Bahman Zamani}, {and} \bibinfo{person}{Shekoufeh
  {Kolahdouz-Rahimi}}.} \bibinfo{year}{2018}\natexlab{}.
\newblock \showarticletitle{{{EVL}}+{{Strace}}: A Novel Bidirectional Model
  Transformation Approach}.
\newblock \bibinfo{journal}{\emph{Information and Software Technology}}
  \bibinfo{volume}{100} (\bibinfo{date}{Aug.} \bibinfo{year}{2018}),
  \bibinfo{pages}{47--72}.
\newblock
\showISSN{09505849}
\urldef\tempurl%
\url{https://linkinghub.elsevier.com/retrieve/pii/S0950584917300629}
\showURL{%
\tempurl}


\bibitem[\protect\citeauthoryear{Sewell, Myreen, and Klein}{Sewell
  et~al\mbox{.}}{2013}]%
        {sewellTranslationValidationVerified2013}
\bibfield{author}{\bibinfo{person}{Thomas Arthur~Leck Sewell},
  \bibinfo{person}{Magnus~O. Myreen}, {and} \bibinfo{person}{Gerwin Klein}.}
  \bibinfo{year}{2013}\natexlab{}.
\newblock \showarticletitle{Translation Validation for a Verified {{OS}}
  Kernel}. In \bibinfo{booktitle}{\emph{Proceedings of the 34th {{ACM SIGPLAN}}
  Conference on {{Programming}} Language Design and Implementation}}.
  \bibinfo{pages}{471--482}.
\newblock


\bibitem[\protect\citeauthoryear{Srivastava, Gulwani, and Foster}{Srivastava
  et~al\mbox{.}}{2010}]%
        {srivastavaProgramVerificationProgram2010}
\bibfield{author}{\bibinfo{person}{Saurabh Srivastava}, \bibinfo{person}{Sumit
  Gulwani}, {and} \bibinfo{person}{Jeffrey~S. Foster}.}
  \bibinfo{year}{2010}\natexlab{}.
\newblock \showarticletitle{From Program Verification to Program Synthesis}. In
  \bibinfo{booktitle}{\emph{Proceedings of the 37th Annual {{ACM
  SIGPLAN-SIGACT}} Symposium on {{Principles}} of Programming Languages}}.
  \bibinfo{pages}{313--326}.
\newblock


\bibitem[\protect\citeauthoryear{Weidmann, Anjorin, Varro, Fritsche, Schurr,
  and Leblebici}{Weidmann et~al\mbox{.}}{2019}]%
        {weidmannIncrementalBidirectionalModel2019b}
\bibfield{author}{\bibinfo{person}{Nils Weidmann}, \bibinfo{person}{Anthony
  Anjorin}, \bibinfo{person}{Gergely Varro}, \bibinfo{person}{Lars Fritsche},
  \bibinfo{person}{Andy Schurr}, {and} \bibinfo{person}{Erhan Leblebici}.}
  \bibinfo{year}{2019}\natexlab{}.
\newblock \showarticletitle{Incremental {{Bidirectional Model Transformation}}
  with {{eMoflon}}::{{IBeX}}}. In \bibinfo{booktitle}{\emph{Proceedings of the
  {{Eighth International Workshop}} on {{Bidirectional Transformations}}}}.
\newblock
\urldef\tempurl%
\url{http://ceur-ws.org}
\showURL{%
\tempurl}


\bibitem[\protect\citeauthoryear{Xu, Wang, Zhan, Jin, Talpin, and Zhan}{Xu
  et~al\mbox{.}}{2022}]%
        {xuUnifiedGraphicalComodeling2022}
\bibfield{author}{\bibinfo{person}{Xiong Xu}, \bibinfo{person}{Shuling Wang},
  \bibinfo{person}{Bohua Zhan}, \bibinfo{person}{Xiangyu Jin},
  \bibinfo{person}{Jean-Pierre Talpin}, {and} \bibinfo{person}{Naijun Zhan}.}
  \bibinfo{year}{2022}\natexlab{}.
\newblock \showarticletitle{Unified Graphical Co-Modeling, Analysis and
  Verification of Cyber-Physical Systems by Combining {{AADL}} and
  {{Simulink}}/{{Stateflow}}}.
\newblock \bibinfo{journal}{\emph{Theoretical computer science}}
  \bibinfo{volume}{903} (\bibinfo{year}{2022}), \bibinfo{pages}{1--25}.
\newblock


\bibitem[\protect\citeauthoryear{Yamaguchi, Matsuda, David, and Wang}{Yamaguchi
  et~al\mbox{.}}{2021}]%
        {yamaguchiSynbitSynthesizingBidirectional2021}
\bibfield{author}{\bibinfo{person}{Masaomi Yamaguchi},
  \bibinfo{person}{Kazutaka Matsuda}, \bibinfo{person}{Cristina David}, {and}
  \bibinfo{person}{Meng Wang}.} \bibinfo{year}{2021}\natexlab{}.
\newblock \bibinfo{title}{Synbit: {{Synthesizing Bidirectional Programs}} Using
  {{Unidirectional Sketches}}}.
\newblock
\newblock
\showeprint[arxiv]{2108.13783}~[cs]
\urldef\tempurl%
\url{http://arxiv.org/abs/2108.13783}
\showURL{%
\tempurl}


\bibitem[\protect\citeauthoryear{Zhan, Gu, Xu, Jin, Wang, Xue, Li, Chen, Yang,
  and Zhan}{Zhan et~al\mbox{.}}{2021}]%
        {zhanBriefIndustryPaper2021}
\bibfield{author}{\bibinfo{person}{Bohua Zhan}, \bibinfo{person}{Bin Gu},
  \bibinfo{person}{Xiong Xu}, \bibinfo{person}{Xiangyu Jin},
  \bibinfo{person}{Shuling Wang}, \bibinfo{person}{Bai Xue},
  \bibinfo{person}{Xiaofeng Li}, \bibinfo{person}{Yao Chen},
  \bibinfo{person}{Mengfei Yang}, {and} \bibinfo{person}{Naijun Zhan}.}
  \bibinfo{year}{2021}\natexlab{}.
\newblock \showarticletitle{Brief Industry Paper: {{Modeling}} and Verification
  of Descent Guidance Control of {{Mars}} Lander}. In
  \bibinfo{booktitle}{\emph{2021 {{IEEE}} 27th {{Real-Time}} and {{Embedded
  Technology}} and {{Applications Symposium}} ({{RTAS}})}}.
  \bibinfo{publisher}{{IEEE}}, \bibinfo{pages}{457--460}.
\newblock


\bibitem[\protect\citeauthoryear{Zhu, Ko, Martins, Saraiva, and Hu}{Zhu
  et~al\mbox{.}}{2015}]%
        {zhuBiYaccRollYour2015}
\bibfield{author}{\bibinfo{person}{Zirun Zhu}, \bibinfo{person}{Hsiang-Shang
  Ko}, \bibinfo{person}{Pedro Miguel~Ribeiro Martins},
  \bibinfo{person}{Jo{\~a}o~Alexandre Saraiva}, {and}
  \bibinfo{person}{Zhenjiang Hu}.} \bibinfo{year}{2015}\natexlab{}.
\newblock \showarticletitle{{{BiYacc}}: {{Roll}} Your Parser and Reflective
  Printer into One}. In \bibinfo{booktitle}{\emph{Proceedings of the {{Fourth
  International Workshop}} on {{Bidirectional Transformations}}}}.
  \bibinfo{publisher}{{CEUR-Ws}}.
\newblock
\urldef\tempurl%
\url{http://ceur-ws.org}
\showURL{%
\tempurl}


\bibitem[\protect\citeauthoryear{Zhu, Ko, Zhang, Martins, Saraiva, and Hu}{Zhu
  et~al\mbox{.}}{2020}]%
        {zhuUnifyingParsingReflective2020a}
\bibfield{author}{\bibinfo{person}{Zirun Zhu}, \bibinfo{person}{Hsiang-Shang
  Ko}, \bibinfo{person}{Yongzhe Zhang}, \bibinfo{person}{Pedro Martins},
  \bibinfo{person}{Jo{\~a}o Saraiva}, {and} \bibinfo{person}{Zhenjiang Hu}.}
  \bibinfo{year}{2020}\natexlab{}.
\newblock \showarticletitle{Unifying {{Parsing}} and {{Reflective Printing}}
  for {{Fully Disambiguated Grammars}}}.
\newblock \bibinfo{journal}{\emph{New Generation Computing}}
  \bibinfo{volume}{38}, \bibinfo{number}{3} (\bibinfo{date}{July}
  \bibinfo{year}{2020}), \bibinfo{pages}{423--476}.
\newblock
\showISSN{1882-7055}
\urldef\tempurl%
\url{https://doi.org/10.1007/s00354-019-00082-y}
\showURL{%
\tempurl}


\bibitem[\protect\citeauthoryear{Zou, Zhan, Wang, and Fr{\"a}nzle}{Zou
  et~al\mbox{.}}{2015}]%
        {zouFormalVerificationSimulink2015}
\bibfield{author}{\bibinfo{person}{Liang Zou}, \bibinfo{person}{Naijun Zhan},
  \bibinfo{person}{Shuling Wang}, {and} \bibinfo{person}{Martin Fr{\"a}nzle}.}
  \bibinfo{year}{2015}\natexlab{}.
\newblock \showarticletitle{Formal Verification of {{Simulink}}/{{Stateflow}}
  Diagrams}. In \bibinfo{booktitle}{\emph{Automated {{Technology}} for
  {{Verification}} and {{Analysis}}: 13th {{International Symposium}}, {{ATVA}}
  2015, {{Shanghai}}, {{China}}, {{October}} 12-15, 2015, {{Proceedings}} 13}}.
  \bibinfo{publisher}{{Springer}}, \bibinfo{pages}{464--481}.
\newblock


\bibitem[\protect\citeauthoryear{Zou, Zhany, Wang, Fr{\"a}nzle, and Qin}{Zou
  et~al\mbox{.}}{2013}]%
        {zouVerifyingSimulinkDiagrams2013}
\bibfield{author}{\bibinfo{person}{Liang Zou}, \bibinfo{person}{Naijun Zhany},
  \bibinfo{person}{Shuling Wang}, \bibinfo{person}{Martin Fr{\"a}nzle}, {and}
  \bibinfo{person}{Shengchao Qin}.} \bibinfo{year}{2013}\natexlab{}.
\newblock \showarticletitle{Verifying Simulink Diagrams via a Hybrid Hoare
  Logic Prover}. In \bibinfo{booktitle}{\emph{2013 {{Proceedings}} of the
  {{International Conference}} on {{Embedded Software}} ({{EMSOFT}})}}.
  \bibinfo{publisher}{{IEEE}}, \bibinfo{pages}{1--10}.
\newblock


\end{thebibliography}

\end{document}